\documentclass[12pt, leqno]{article}
\usepackage{amsmath,mathrsfs,bm,amssymb,color}

\newtheorem{theorem}{Theorem}[section]
\newtheorem{proposition}[theorem]{Proposition}
\newtheorem{lemma}[theorem]{Lemma}
\newtheorem{corollary}[theorem]{Corollary}
\newtheorem{definition}[theorem]{Definition}
\newtheorem{example}[theorem]{Example}
\newtheorem{remark}[theorem]{Remark}

\newcommand{\eps}{\epsilon}

\newcommand{\N}{{C_N}}
\newcommand{\W}{{\rm W}}
\newcommand{\UU}{\ms U}
\newcommand{\GG}{{\mathcal G}}
\newcommand{\PPPP}{{\rm P}}

\newcommand{\proof}{{\noindent \it Proof:\ }}
\newcommand{\qed}{\hfill $\blacksquare$\par\medskip}
\newcommand{\BR} {{\mathbb {R}^d }}
\newcommand{\BRN}{{\mathbb {R}^{dN} }}

\newcommand{\CC}{{\mathbb{C}}}
\newcommand{\RR}{\mathbb{R}}
\newcommand{\hp}{H_{\rm p}}
\newcommand{\eff}{V_{{\rm eff}}}
\newcommand{\heff}{h_{\rm eff}}
\newcommand{\weff}{W_{\rm eff}}

\newcommand{\e}{\epsilon}
\newcommand{\QQQ}{\ms Q}

\def\bbbone{{\mathchoice {\rm 1\mskip-4mu l} {\rm 1\mskip-4mu l}
{\rm 1\mskip-4.5mu l} {\rm 1\mskip-5mu l}}}
\def\one{\bbbone}

\newcommand{\bx}{{\bf x}}
\newcommand{\BX}{{\bf X}}
\newcommand{\bz}{{\bf 0}}

\newcommand{\pt}{P_{\rm tot}}
\newcommand{\ps}{\sum_{j=1}^N p_j}
\newcommand{\Ebb}{{\mathbb E}}
\newcommand{\dG}{{\rm d}\Gamma}
\newcommand{\Hr}{H_{\rm R}}
\newcommand{\cut}{\chi}

\newcommand{\LR}{{L^2(\BR)}}
\newcommand{\LRT}{{L^2(\RR^{dN})}}

\newcommand{\pro}[1]{(#1_t)_{t\geq0}}

\newcommand{\fff}{{\mathscr F}}

\newcommand{\ms}[1]{{\mathscr #1}}
\newcommand{\lk}{\left(}
\newcommand{\rk}{\right)}

\newcommand{\ov}[1]{\overline{#1}}
\newcommand{\hf}{H_{\rm f}}
\newcommand{\hfs}{H_{{\rm f},\s}}

\newcommand{\gr}{\Phi_\sigma}
\newcommand{\grp}{\phi_{\rm p}}

\newcommand{\half}{\frac{1}{2}}
\newcommand{\han}{{1/2}}
\newcommand{\hi}{H_{\rm I}}
\newcommand{\ki}{K_{\rm I}}
\newcommand{\his}{H_{{\rm I},\s}}

\newcommand{\la}{\hat \lambda}
\newcommand{\s}{\sigma}
\newcommand{\hhh}{{\ms H}}

\renewcommand{\d}{\displaystyle}
\newcommand{\non}{\nonumber}

\newcommand{\BBB}{{\ms B}}

\newcommand{\atopb}[2]{\genfrac{}{}{0pt}{}{#1}{#2}}
\newcommand{\cE}{\mathcal{E}}

\def\til#1{\widetilde{#1}}

\newcommand{\vep}{\varepsilon}
\newcommand{\ome}{\omega}

\newcommand{\cW}{\W}
\newcommand{\cK}{\mathcal{K}}

\newcommand{\ck}{\mathcal{K}}
\newcommand{\tensor}{\otimes}
\newcommand{\inner}[2]{\lk #1,#2\rk }
\newcommand{\norm}[1]{ \left\| {#1} \right\| }

\newcommand{\Ran}{\mathrm{Ran}}

\makeatletter
    
    \@addtoreset{equation}{section}
  \makeatother

\title{\sc Enhanced binding of an $N$-particle system interacting with a scalar field II.Relativistic version}

\author{Fumio Hiroshima\\ 
{\small\it Faculty of Mathematics, Kyushu University}\\
{\small\it 744 Motooka, Fukuoka, 819-0395, Japan}   \\
{\small {\sf hiroshima@math.kyushu-u.ac.jp}}\\ \ \\
Itaru Sasaki \\ 
{\small\it Department of Mathematics, Shinshu University}\\
{\small \it 3-1-1 Asahi, Matsumoto, 390-8621, Japan} \\
{\small {\sf isasaki@shinshu-u.ac.jp}} \\ 
}
\date{}

\begin{document}
\setlength{\baselineskip}{15pt}
\maketitle

\begin{abstract}
An enhanced binding of $N$-{\it relativistic} particles coupled 
to a massless scalar bose field is investigated.
It is not assumed that the system has a ground state for the zero-coupling.
It is shown, however, that there exists a ground state for sufficiently large coupling.
The proof is based on checking the stability condition and showing
a uniform exponential decay of infrared regularized ground states.
\end{abstract}
\section{Preliminaries}
\subsection{Introduction}
Non-perturbative analysis of eigenvalues embedded in the continuous
 spectrum has been developed in the last decade and 
 it has been applied to the
 mathematically rigorous analysis of the spectra of self-adjoint Hamiltonians in quantum field theory.
Among other things, stability and instability of a quantum mechanical particle
 coupled to a quantum field have been investigated.

The Hamiltonian in quantum field theory is realized as a self-adjoint operator 
of the form 
\begin{align}
\label{K}
K_0+\alpha \ki,
\end{align}
acting in a Hilbert space over $\CC$ for each values of coupling constant $\alpha\in\RR$.  Here $K_0$ is the subject term and $\ki$ an interaction term.
We are concerned with ground states of $K_0+\alpha\ki$ in this paper. 

Let $\s(T)$ be the spectrum of a self-adjoint operator $T$. 
\begin{definition}
\label{IIII}
{\rm ({\bf Ground state and ground state energy}) 
 Let $T$ be a self-adjoint operator bounded from below. Then 
the bottom of the spectrum, $E_0(T)=\inf\sigma(T)$,  is called a ground state energy of $T$.
Let  $E_0(T)$ be  an eigenvalue of $T$. Then 
the  eigenvector $f$ associated with $E_0(T)$ is is called a
ground state of $T$, i.e., $Tf=E_0(T) f$.
}\end{definition}
Generally  the bottom of the spectrum of the zero-coupling Hamiltonian $K_0$
coincides with the bottom of the continuous spectrum of $K_0$. 
Then  the bottom is embedded in the continuum and in particular it is emphasized not to be discrete. 
Hence the spectral analysis of $K_0+\alpha \ki$ is regarded as 
the perturbation problem of embedded eigenvalues. 
Although an analytic perturbation theory of the discrete spectrum is established for a various type of self-adjoint operators, 
the perturbation of embedded eigenvalues are crucial and it is not straightforward to apply the perturbation theory of discrete spectra. 
Then it is subtle to show the existence of a ground state  of $K_0+\alpha\ki$ 
not only for arbitrary values of coupling constant but also small values of coupling constant. 
Moreover it is not necessarily  that  a ground state exists  for 
$K_0+\alpha \ki$, $\alpha\ne 0$, 
even when $\inf \s(K_0+\alpha \ki)>-\infty$  and  $K_0$ has  ground state.

The existence and the absence of a ground state for physically reasonable Hamiltonians of quantum field theory 
has  been however proven so far under some assumptions.  
The existence of the ground state of the standard Nelson Hamiltonian \cite{nel64} 
was in particular proven in e.g.,
\cite{bfs98,spo98,ger00,sas05}, 
where the most basic assumptions for proving the existence of a ground state 
are 
\begin{description}
\item[(1)]
 infrared regular condition,
 \item[(2)]
 the existence of ground state of $K_0$. 
\end{description}
In particular assumption (2) tells us that  Hamiltonians $K_0+\alpha\ki $  \emph{also} has a ground state for arbitrary values of $\alpha$. 

It is  found however that an interaction with  quantum fields  enhances the binding energy, 
which suggests that a Hamiltonian with \emph{sufficiently large}  coupling constants may have a ground state whether $K_0$  has a ground state or not. 
If $K_0+\alpha\ki $ with  {sufficiently large}  coupling constants has  a ground state 
whether $K_0$ has a ground state or not, then it is said 
that \emph{enhanced binding} occurs. 
Enhanced binding is initiated by \cite{hs01} and in the previous paper \cite{hs08} enhanced binding is shown for a  system of
 $N$-\emph{nonrelativistic} particles governed by Sch\"odinger operator and linearly coupled to a massless scalar bose field.
In this paper replacing the nonrelativistic particles with \emph{relativistic} ones, 
we show
the enhanced binding. 


Finally we give some comments on related works on enhanced binding.
The enhanced binding is studied so far for the various kind of models
 in quantum field theory.
In \cite{hs01} the enhanced binding of the Pauli-Fierz model with the dipole
 approximation is studied. In \cite{hss11} a complement result of \cite{hs01}
 is established, i.e., the absence of ground state for sufficiently small 
 coupling constant is shown. 
 See also \cite{ak03,blv05,bv04,ceh03,cvv03,hvv03} for the related works.

\subsection{Main results}
The total Hamiltonian we consider is of the form
\begin{align} \label{r1}
H^V &=H_0+\kappa \hi.
\end{align}
The operator $H_0=H_0(\kappa)$ describes the zero coupling Hamiltonian and is given by
\begin{align*}
H_0 &= \hp
 + \kappa^2 \hf,\\
 \hp&=\sum_{j=1}^N
\lk \sqrt{-\Delta_j+m_j^2}-m_j + V(x_j) \rk, 
\end{align*}
where $m_j>0$ is the mass of the $j$-th particle, $V(x)$ an external potential,
$\hf$ the free field Hamiltonian, and $\kappa>0$ denotes a scaling parameter.
The operator $H_I$ describes a particle-boson linear interaction.
 We notice that there are no pair potentials in $H^V$ and $V$ is assumed to be
 independent of $j$ for simplicity.
Introducing a dressing transformation $e^{iT}$ to derive an effective potential $\eff$,
we transform $H^V$ as
\begin{align} \label{t1}
 e^{-iT} H^V e^{iT} = \heff^V+ \kappa^2 \hf+H_{\rm R}(\kappa),
\end{align}
where $\heff^V $ is the effective particle Hamiltonian given by
\begin{align}{\label{yu1}}
   \heff^V = \sum_{j=1}^N \lk \sqrt{-\Delta_j+m_j^2} - m_j + V(x_j) \rk + \eff(x_1,...,x_N)
\end{align}
and $\Hr(\kappa)$ a remainder term to be regarded as a perturbation of $\heff^V+\kappa^2\hf$.
Compensating for deriving $\eff$ through the dressing transformation, we have the remainder term $\Hr(\kappa)$ which is unfortunately no longer linear and is the complicated form:
\[
\Hr(\kappa)=\sum_{j=1}^N \lk
\sqrt{\lk -i\nabla_j-\frac{1}{\kappa} A_j(x_j)\rk^2+m_j^2}
 - \sqrt{-\Delta_j+m_j^2}\rk,
\]
where $A_j$ denotes some quantum vector field.
Nevertheless it turns to be a {\it small} perturbation for sufficiently
 large $\kappa$ in some sense.

We are interested in the existence of a ground state of $H^V$,
 equivalently that of $e^{-iT} H^V e^{iT}$.
We do not however assume the existence of ground states of $H_0$.
As will be shown below, the enhanced binding is observed 
by the transformed Hamiltonian \eqref{t1} rather than $H^V$ itself.
Since we consider a massless boson, the bottom of the spectrum of $H^V$
is the edge of the continuous spectrum and the regular perturbation
can not be  applied.
Then it is not clear whether $e^{-iT} H^V e^{iT} $ has a ground state 
even when $\heff^V$ has a ground state.

The conventional approach is to assume an infrared cutoff in the form factor $\la$ in $\hi$
by setting $\la(k)\lceil_{|k|>\s}$,  $H^V$ with cutoff $\la\lceil_{|k|>\s}$ is denoted
 by $H_\s^V$, and to show the existence of a ground state $\gr$ of $H_\s^V$.
The vector $\gr$ is called an infrared-regularized ground state.
Then one is left to show that the sequence of ground states $\gr$ has a non-zero
weak limit $\Phi$ as $\s\to 0$, which is the desired ground state of $ H^V $.
We show in this paper:
\begin{description}
\item[(A)]
 the stability condition for $H^V$ is satisfied (Lemma \ref{P4.2}),
\item[(B)]
  infrared-regularized ground states $\gr$ has exponential decay uniformly with respect to the infrared cutoff parameter $\s$
  (Lemma \ref{expdecay}),
\item[(C)]
we prove that (1) stability condition and (2) exponential decay 
imply the existence of a ground state of $H^V$ (Appendix \ref{appa}), 
\item[(D)]
 we show that there exist $\bar \alpha>0$ and $\kappa_0$ such that 
 for each  $\kappa>\kappa_0$,
 $H^V$ has the unique 
 ground state 
for  
 $|\alpha|\in (\bar \alpha, \bar \alpha(\kappa))$ with some $\bar \alpha(\kappa)$
(Theorem \ref{enhanced}).
\end{description}
Statement (D) describes the enhanced binding and this is the main theorem in this paper.

\subsection{Strategies}
We explain more technical improvement of this paper.

{\bf (Reduction to the stability condition of $\heff^V$)}
 The stability condition is introduced in \cite{gll01} to show the existence
 of ground state of the non-relativistic quantum electrodynamics.
 The key ingredient in this paper is that we reduce the stability condition of $H^V$
 to that of $\heff^V $ in Lemma \ref{P4.2}.
Namely we show that the stability condition of $\heff^V $ implies that of $H^V$.
These are proven by energy comparison inequality derived by functional integration
 of the heat semigroup generated by \eqref{t1}
(Lemma \ref{enebound1}) and a simple variation principle (Lemma \ref{enebound2}),
hence we focus on showing the stability condition on $\heff^V $ instead of $H^V$.

{\bf (Uniform exponential localization by functional integrations)}
Our method is a minor but nontrivial modification of \cite{hs08} and a mixture of
\cite{ger00,gll01}. 
We do not assume the compactness condition on $\hp$, 
which is entered in \cite{ger00}.
Instead of this we show an exponential localization of infrared-regularized ground states,
$\|\gr(x)\|_\fff\leq C_\delta e^{-\delta|x|}$, which is
derived through functional integrations in Lemma \ref{expdecay}.
The crucial point is to show that this localization is uniform on $\sigma>0$, i.e.,
$C_\delta$ and $\delta$ are independent of $\s>0$.

{\bf (Scaling parameter)}
The scaling parameter introduced in this paper can be obtained by replacing 
the annihilation operator $a$ and the creation operator $a^{\ast}$ with 
$\kappa a$ and $\kappa a^{\ast}$, respectively.
This scaling is introduced in 
 \cite{dav77,dav79} and 
 the scaling limit as $\kappa\to\infty$ is called the weak coupling limit. 
Roughly speaking at least in the nonrelativistic domain $\hp\cong -\frac{1}{2m}\Delta+V$,
 and then
\begin{equation*}
 H^V=\kappa^2(\kappa^{-2}\hp+\hf+\kappa^{-1} \hi)
\end{equation*}
with
\begin{equation*}
 \kappa^{-2}\hp\cong -\frac{1}{2m\kappa^2}\Delta+\frac{1}{\kappa^2}V.
\end{equation*}
Thus we interpret  that enhanced binding of $H^V$ occurs
 when sufficiently heavy particle mass and shallow external potential are assumed.
Alternate explanation of the scaling parameter is the tool to derive a Markov process
 from $e^{-tH^V}$.
Although the scalar product $(f\otimes \Omega, e^{-tH^V}g\otimes\Omega)$ does not define a Markov process,
$(f,e^{-t(\heff-E_\mathrm{diag})}g)$ does with generator $\heff-E_\mathrm{diag}$.
This can be obtained by the scaling limit:
$$(f\otimes \Omega, e^{-tH^V}g\otimes\Omega)\to(f,e^{-t(\heff-E_\mathrm{diag})}g)$$ as $\kappa\to\infty$.
More precisely if $\heff$ has a unique positive ground state $\grp$, 
then there exists a Markov process $\pro Y$ such that 
$$(f\grp,e^{-t(\heff-E_\mathrm{diag})}g\grp)=\Ebb[f(Y_0) g(Y_t)],$$ 
where $\Ebb$ denotes the expectation.  


The organization of this paper is as follows.

In the remainder of Section 1 we define the Nelson model with a relativistic kinetic term, 
and introduce a scaling parameter $\kappa>0$.
In Section 2 we introduce a dressing transformation, and mention the stability condition
 and uniform exponential decay of $\gr(x)$.
In Section 3 we prove the stability condition in Section 3.1
and uniform exponential decay in Section 3.2, and in Section 3.3 we show the enhanced binding.

In Appendix \ref{appa} we show that
the relativistic version of the stability condition also implies the existence of the ground state.
In Appendix \ref{ess} we review fundamental properties of the bottom of the essential
spectrum of relativistic Schr\"odinger operator.
In Appendix \ref{appb} we give the functional integral representation of $e^{-tH^V}$
and show some inequality used in the proof of exponential decay of infrared regularized ground states.
In Appendix \ref{appc} we derive some energy comparison inequality of
the translation invariant Hamiltonian
 $\sum_{j=1}^N (\sqrt{-\Delta_j+m_j}-m_j+V(x_j))$.

\subsection{Definition}
 We begin with giving the definition of the Nelson model with $N$-relativistic particles.
Throughout we assume $N\geq 2$ and the dimension of state space is $d\geq 3$.
The Hamiltonian of the Nelson model can be realized as a self-adjoint operator on the tensor product of $\LRT$
and the boson Fock space $\fff$ over $\LR$,
 \begin{align} \label{r2}
 \hhh=\LRT\otimes\fff.
 \end{align}
 Here $\fff$ is defined by $\fff=\oplus_{n=0}^\infty L_{\rm sym}^2(\RR^{dn})$,
where $L_{\rm sym}^2(\RR^{dn})$ is the set of square integrable functions such that
$\Psi(x_1,\cdots,x_n)=\Psi(x_{\sigma(1)},\cdots,x_{\sigma(n)})$ for any $n$-degree permutation $\sigma$.
A vector $\Psi\in \fff$ is written as $\Psi=\{\Psi^{(n)}\}_{n=0}^\infty$
with $\Psi^{(n)}\in L_{\rm sym}^2(\RR^{dn})$, and the Fock vacuum $\Omega\in\fff$ is defined by
$\Omega=\{1,0,0,\ldots\}$.
We denote by $a(f)$ and $a^{\ast}(f)$, $f\in\LR$, the annihilation and creation operator in $\fff$, respectively.
 They satisfy canonical commutation relations:
\begin{align}
 [a(f), a^{\ast}(g)]=(\bar f, g)\one ,\ \ \ [a(f), a(g)]=0=[a^{\ast}(f),a^{\ast}(g)]
\end{align}
and the adjoint relation $a^{\ast}(f)=(a(\bar f))^\ast$ holds.
Throughout this paper $(F,G)_{\ck}$ denotes the scalar product on Hilbert space $\cK$, which is linear in $G$ and antilinear in $F$.
We omit $\cK$ until confusions arises. We informally write as
$ a^\#(f)=\int a^\#(k) f(k) dk$, $a^\#=a, a^{\ast}$.
The second quantization
of the closed operator $A$ on $\LR$ is denoted by $\dG (A)$.
The free field Hamiltonian $\hf$ is the self-adjoint operator on $\fff$,
which is given by the second quantization of the multiplication operator
 $\omega(k)=|k|$ on $\LR$:
\begin{align}\label{3}
\hf = \dG (\ome).
\end{align}
Next we introduce particle Hamiltonian.
We suppose that the $N$-relativistic particles are governed by
the relativistic Schr\"odinger operator $\hp$ of the form:
\begin{align} \label{1}
  \hp=\sum_{j=1}^N (\Omega_j+V_j)
\end{align}
which is acting on $\LRT$, where
\begin{align} \label{2}
  \Omega_j=\Omega_j(p_j )=\sqrt{p_j ^2+m_j^2}-m_j,
\end{align}
is the $j$-th particle Hamiltonian with momentum $p_j =-i\nabla_{x_j}$ and mass $m_j>0$.
$V_j=V(x_j)$ denotes an external potential.
In this paper, we assume that there is no interparticle potential for simplicity.

The Hamiltonian of the relativistic Nelson model is then defined by
\begin{align} \label{10}
  H^V = H_0+\kappa \hi,
\end{align}
where the zero-coupling Hamiltonian $H_0$ is given by
\begin{align} \label{11}
  H_0=\hp\otimes \one + \kappa^2 \one \otimes \hf
\end{align}
and $\kappa>0$ a scaling parameter.
$\hi$ denotes the linear interaction given by
\begin{align} \label{6}
 \hi= \alpha\sum_{j=1}^N \int_{\RR^{dN}}^\oplus \phi_j(x_j) dX
\end{align}
under the identification:
$\d \hhh \cong \int^\oplus_{\RR^{dN}} \fff dX$,
where $dX = dx_1\cdots dx_N$.
Here $\alpha\geq 0$ is a coupling constant, and the scalar field $\phi_j(x)$ is given by
\begin{align} \label{7}
\phi_j(x) = \overline{\frac{1}{\sqrt 2}\int_\BR (a^{\ast}(k) \la _j(-k) e^{-ikx}+a(k) \la _j(k) e^{ikx} ) dk}
\end{align}
for each $x\in\BR$ with ultraviolet cutoff functions $\la _j$.
Here $\ov{\{\cdots\}}$ denotes the operator closure. 
The standard choice of the ultraviolet cutoff is $\la_j(k) = (2\pi)^{-d/2}\omega(k)^{-1} \one_{|k|\leq \Lambda}$,
where $\one_X$ denotes the characteristic function of $X$.
We do not however fix any special cutoff function.

Throughout this paper we assume the following three conditions:
\begin{description}
\item[(V)] $V(-\Delta+1)^{-1/2}$ is compact.
\item[(UV)]
$\la _j(-k)={\la _j(k)}\geq 0$ and $\la _j\in\LR$ for $j=1,...,N$.
\item[(IR)] $\la _j/\omega\in\LR$ for $j=1,...,N$.
\end{description}
Assumption (V) implies that $V$ is infinitesimally small with respect to $\sqrt{-\Delta+m^2}-m$
for all $m\geq 0$.
Hence, by the Kato-Rellich theorem, $H_\mathrm{p}$ is self-adjoint on
$D(\sum_{j=1}^N\Omega_j)$ and essentially self-adjoint on any core for $\sum_{j=1}^N\Omega_j$,
where $D(A)$ denotes the domain of $A$.
(UV) implies that $\hi$ is symmetric.
Then (V), (UV) and (IR) also imply that, for arbitrary $\alpha\in\RR$ and $\epsilon>0$,
it holds that
\begin{align*}
 \|\hi\Psi\|\leq \epsilon \|H_0\Psi\|+b_\epsilon
 \|\Psi\|, ~~~~ \Psi\in D(H_0).
\end{align*}
Therefore, by the Kato-Rellich theorem, $H^V$ is self-adjoint on $D(H_0)$ for all $\kappa>0$ and $\alpha\geq 0$.
The nonnegativity $\la_j(k)\geq 0$ in (UV) implies that the effective potential is attractive, which is used in Lemma \ref{ene}.

\section{Existence of a ground state}
\subsection{Dressing transformation}
To derive the effective particle Hamiltonian we introduce the so-called dressing transformation
$e^{-iT}$, where $T=\frac{\alpha}{\kappa}\sum_{j=1}^N \pi_j$ and
\begin{equation*}
   \pi_j = \int_{\RR^{dN}}^\oplus dX
   \lk \overline{ \frac{i}{\sqrt2} \int \lk a^{\ast}(k) e^{-ikx_j} \frac{\la _j(-k)}{\omega(k)} - a(k) e^{ikx_j} 
   \frac{\la _j(k)}{\omega(k)} \rk dk} \rk.
\end{equation*}
By (IR), $\pi_j$ is self-adjoint on $\hhh$ and then $e^{iT}$ is unitary.
\begin{lemma} \label{P4.1}
The unitary operator $e^{iT}$ maps $D(H^V)$ onto itself and
\begin{align} \label{sa2}
 e^{-iT} H^V e^{iT}
 = \heff^V \otimes \one+\kappa^2 \one \otimes \hf+\Hr(\kappa),
\end{align}
where the effective Hamiltonian is defined by
\begin{align}
   \heff^V = \sum_{j=1}^N (\Omega_j+V_j) + \eff,
\end{align}
with the effective pair potential
\begin{align}
   \eff &= \alpha^2 \sum_{1\leq i<j\leq N} W_{ij}(x_i-x_j),\\
   W_{ij}(x) &= - \int_{\BR} \frac{\hat{\lambda}_i(-k)\la _j(k)}{\omega(k)} e^{-ikx} dk.
\end{align}
Here $\Hr(\kappa)$ is the remainder term given by
\begin{align}
   \Hr(\kappa) & = \sum_{j=1}^N \lk
   \Delta\Omega_j - \frac{\alpha^2}{2} \|\la _j/\sqrt\omega\|^2 \rk,\\
   \Delta\Omega_j&=\Omega_j\lk p_j +\frac{\alpha}{\kappa}A_j \rk
   - \Omega_j\lk p_j \rk
\end{align}
with a vector field
\begin{align*}
  & A_j = (A_{j1},\cdots,A_{jd}) \\
  & A_{jl} = \int_{\RR^{dN}}^\oplus\Bigg(\frac{1}{\sqrt 2} \overline{ \int_\BR k_l \lk a^{\ast}(k) e^{-ikx_j}
\frac{\la _j(-k)}{\omega(k)} + a(k) e^{ikx_j} \frac{\la_j(k)}{\omega(k)}\rk dk }\Bigg) dX.
\end{align*}

\end{lemma}
\proof
We directly see that
\begin{align*}
e^{-iT} p_j e^{iT}& = p_j+\frac{\alpha}{\kappa}A_j,\\
e^{-iT} \phi_j e^{iT} &= \phi_j - \frac{\alpha}{\kappa} \sum_{i=1}^N
 \int_\BR \frac{\hat\lambda_i(k)\hat\lambda_j(-k)}{\omega(k)}e^{-ik(x_j-x_i)}dk, \\
 e^{-iT} \hf e^{iT} &= \hf - \frac{1}{\kappa} \hi +
  \frac{\alpha^2}{2\kappa^2} \sum_{i,j=1}^N \int_{\BR}
 \frac{\hat\lambda_i(-k)\hat\lambda_j(k)}{\omega(k)}e^{-ik(x_i-x_j)}dk.
\end{align*}
Together with them, the lemma follows.
\qed
(UV) and (IR) imply that $\eff$ is bounded.
Therefore $H^V_{\rm eff}$ is a self-adjoint operator on $D(\sum_{j=1}^N\Omega_j)$.

\subsection{Main results}
Recall that $E_0(T)=\inf\s (T)$ for a self-adjoint operator $T$.  
\begin{theorem}{\bf (Existence of ground state)} \label{main}
Assume (V), (UV) and (IR).
Suppose that 
$E_0(h^V_{\rm eff}) \in \sigma_{\rm disc}(h^V_{\rm eff})$.
Then there exists $\kappa_0>0$ such that  
$H^V$ has the unique ground state  
for  any $\kappa>\kappa_0$. 
\end{theorem}
In order to show the enhanced binding, we introduce an assumption on $V$.
\begin{itemize}
 \item[{ (EN)}]
 \begin{itemize}
 \item[(1)]
 $\d \inf_{x\in\BR} V(x)>-\infty$ and $\d\liminf_{|x|\to\infty} V(x)=0$;
\item[ (2)]
 $\sqrt{-\Delta}+N V$ acting in $\LR$ has a negative energy ground state;
\item[(3)] $V$ is $d$-dimensional relativistic Kato-class, i.e.,
$$\lim_{t\downarrow 0}\sup_{x\in\BR}\Ebb_\PPPP^x
\left[
\int_0^t V(X_s) ds\right]=0,$$
where $\Ebb_\PPPP^x$ denotes the expectation on
 a probability space $({\cal D},{\cal B},\PPPP^x)$, and
 $\pro X$ denotes the $d$-dimensional L\'evy process
with the characteristic function
$\Ebb_\PPPP^x[e^{iuX_t}]=e^{-t(\sqrt{u^2+m^2}-m)}e^{iux}$.
\end{itemize}
\end{itemize}
Assumption (EN)(1) is used 
only to show spatial exponential decay of the infrared regularized ground state $\gr$. 
The second assumption (EN)(2), which is used in \eqref{yy3}, is a crucial assumption for showing the enhanced binding.
Intuitively a sufficiently strong interaction  engages $N$ particles through linear interaction of the quantum field,  
and consequently  the total Hamiltonian can be regarded as 
$\sqrt{-\Delta} +NV$. 
This intuitive description is  justified in this paper. 
(EN)(3) is used to show the continuity of ground state energy of a translation invariant
 Hamiltonian in Lemma \ref{p=0}.

We state the main results in this paper.

\begin{theorem}{\bf (Enhanced binding)} 
\label{enhanced}
Suppose (V), (UV) and (IR). 
Assume  (EN) and $N\geq 2$.
Then there exist $\bar\alpha>0$ and $\kappa_0>0$ such that 
for each $\kappa>\kappa_0$, $H^V$ has the unique ground state  
for  $|\alpha|\in (\bar \alpha, \bar \alpha(\kappa))$ with some 
constant $\bar\alpha(\kappa)$.
\end{theorem}
\begin{remark} \label{glll}
{\rm In Theorem \ref{main} $\heff^V$ has a ground state. 
In Theorem \ref{enhanced} we do {\it not} assume the existence of a ground state of
$\hp$, i.e., the zero-coupling Hamiltonian $H_0$ does not necessarily have a ground state.  
}\end{remark}
\begin{remark}
{\rm In the case of $N=1$, we can not apply our method to show the enhanced binding. 
Although in this case the enhanced binding may also occur, 
it is crucial to estimate dressing transformed Hamiltonian \eqref{sa2}.   
We do not then discuss this case. 
}\end{remark}
\begin{example}
{\rm We give examples of $V$ satisfying (V) and (EN), but $\sqrt{-\Delta+1}-1+V$ has no ground state
in the dimension $d\geq 3$.
Suppose that $\tilde V$ satisfies 
$$|\tilde{V}(x)|\leq c(1+|x|)^{-\epsilon}$$
with some $c>0$ and $\epsilon>0$. 
It involves $\tilde{V} = -e^{-x^2}$.
 Then (V) is satisfied with $V=\delta\tilde{V}$ for all constant $\delta >0$.
Let $\tilde{V}\not\equiv 0$, $\tilde{V}\leq 0$ and $\tilde{V}\in L^d(\BR)\cap L^{d/2}(\BR)$.
 Let $\delta>0$ be sufficiently small constants and set
\begin{align} \label{hiro}
 H_\delta=\sqrt{-\Delta+1}-1+\delta \tilde{V}.
\end{align}
Let $E_\delta(\cdot)$ be the spectral measure of $H_\delta$.
Since  $\tilde V (\sqrt{-\Delta+1})^{-1}$ is compact,
the essential spectrum of $H_\delta$ is $\s_{\rm ess}(H_\delta)=[0,\infty)$ for all $\delta>0$. 
By the relativistic version of the Lieb-Thirring bound \cite{dau83}, we have
\begin{align}
\label{lt}
 \dim\Ran E_\delta((-\infty,0])
 \leq c_1 \delta^d \int_\BR | \tilde{V}(x)|^d dx + c_2 \delta^{d/2}\int_\BR | \tilde{V}(x)|^{d/2}dx, 
\end{align}
where $c_1$ and $c_2$ are positive constants independent of $\tilde V$.
Hence $H_\delta$ has no ground state for sufficiently small $\delta$ such 
that the right-hand side of \eqref{lt} is strictly smaller than one. 
Similarly $\s_{\rm ess}(\sqrt{-\Delta} +N\delta \tilde V)=[0,\infty)$ follows. 
$\sqrt{-\Delta} +N\delta \tilde V$ has however 
 a negative eigenvalue for sufficiently large $N$, 
 since 
 $\inf \sigma(\sqrt{-\Delta}+N\delta \tilde V)<0$ 
for sufficiently large $N$, which implies that $\sqrt{-\Delta}+N\delta \tilde V$ has a ground state for sufficiently large $N$.
Therefore for sufficiently small $\delta$,  $V=\delta \tilde V$  satisfies (V) and (EN), but 
$\sqrt{-\Delta+1}-1+\delta \tilde{V}$ has no ground state. 
}\end{example}

\subsection{Stability condition and exponential decay}
In order to prove Theorems \ref{main} and \ref{enhanced} we investigate the stability condition.
First of all we introduce cluster Hamiltonians.
Let $C_N=\{1,2,\cdots,N\}$.
For each $\beta \subset C_N,(\beta\neq \emptyset)$,
we define
\begin{align}
 H^0(\beta) &= \sum_{j\in\beta} (\Omega_j+\kappa\alpha \phi_j)+\kappa^2\hf, \\
 H^V(\beta) &= H^0(\beta) + \sum_{j\in\beta} V_j,
\end{align}
acting on $L^2(\mathbb{R}^{d|\beta|})\otimes \fff $, where
$\phi_j = \int_{\RR^{d|\beta|}}^\oplus \phi_j(x_j)dX_\beta$, $X_\beta= (x_j)_{j\in\beta}$.
Clearly $H^V=H^V(C_N)$.
Let
\begin{align} \label{s1}
E^0(\beta)=\inf\sigma(H^0(\beta)),\quad
E^V(\beta)=\inf\sigma(H^V(\beta)).
\end{align}
For the case of $\beta=\emptyset$, we set $E^0(\emptyset )=E^V(\emptyset)=0$.
The lowest two cluster threshold is defined as the minimal energy of systems
 such that only the particles involved in $\beta$ are bound by the origin
but others are sufficiently remote from the origin.
It is defined by
\begin{align}
 \Sigma^V= \min\{E^V(\beta)+E^0(\beta^c)|\beta\subsetneqq C_N \} \label{threshold}
\end{align}
The gap between the ground state energy $E^V$ and the lowest two 
cluster threshold $\Sigma^V$ is related to the existence of ground 
state by the proposition below.
Let $H_\s^V$ be defined by $H^V$ with $\la_j$ replaced by $\la_j(k)\one_{|k|>\s}$.
\begin{proposition}
{\label{P3.1}}
\begin{description}
\item[(Case $\s>0$)]
Suppose that $ E^V < \Sigma^V$.
Then $H_\s^V$ has the unique ground state.
We denote the ground state by $\gr$.
\item[(Case $\s=0$)]
Suppose that $ E^V < \Sigma^V$ and there exists $0<\delta$ independent of $\s$ such that
$\sup_{0<\s<{\bar\s}}
\|(e^{\delta|X|}\otimes\one)\gr\|_\hhh<\infty$ with some ${\bar\s}>0$.
Then $H^V$ has a ground state.
\end{description}
\end{proposition}
\proof
 The proof is a minor modification of \cite{ger00,gll01},
and it is given in Appendix \ref{massive} for the case $\s>0$,
and in Appendix \ref{massless} for the case $\s=0$.
\qed
The condition $\Sigma^V>E^V$ is called the stability condition.
For our model the uniform exponential decay of $\|\gr(x)\|_\fff$ may be derived from the stability condition, but we do not check it. So we need not only stability condition but also uniform exponential decay.

\section{Proof of the main theorem}
In order to show Theorems \ref{main} and \ref{enhanced},
 by Proposition \ref{P3.1} it is enough to show both 
 (1) stability condition and (2) the uniform exponential
 decay of $\|\gr(x)\|_\fff$.
\subsection{Stability condition}
It is however not straightforward to show the stability condition, so we will make a detour and the discussion will be reduced to that of effective particle Hamiltonian $\heff^V$.
Let us define the lowest two cluster threshold of $\heff^V$ in a similar way to $H^V$ and
we shall compare it with
 $\Sigma^V$.
For $\beta \subset C_N$, we define effective cluster Hamiltonians
by \begin{align}
 \heff^0(\beta) & = \sum_{j\in\beta} \Omega_j
 -\alpha^2 \sum_{i,j\in\beta, i<j} W_{ij}(x_i-x_j),\\
 \heff^V(\beta) & = \heff^0(\beta)+\sum_{j\in\beta}V_j.
\end{align}
We set
\begin{align}
 \cE^0(\beta) = \inf\sigma(\heff^0(\beta)), \qquad
 \cE^V(\beta) = \inf\sigma(\heff^V(\beta))
\end{align}
and $\cE^V = \cE^V(C_N)$.
Then the lowest two cluster threshold of $\heff^V$ is defined by
\begin{align}{\label{sa3}}
 \Xi^V
 = \min\{ \cE^V(\beta) + \cE^0(\beta^c)| \beta\subsetneqq C_N\}.
\end{align}
Constants $c^V$ and $d^V$ are such that
$ \|\sum_{j=1}^N \Omega_j \Psi \|
 \leq c^V \|\heff^V \Psi \| + d^V \|\Psi\|$ and set
\begin{align} \label{poisson}
 \GG(t) =
 \lk \sum_{j=1}^N
 \|{\la _j}/{\omega}\|
 \|\la_j\|
 \rk t^2+
 \lk \sum_{j=1}^N
\sqrt 2 m_j
 \|{\la _j}/{\omega}\|
\rk |t|
 + \sqrt{2}
N  \left( c^V |\cE^V| + d^V \right).
  \end{align}
  The next lemma is a key ingredient of this paper.
\begin{lemma}\label{P4.2}
We assume that $\Xi^V-\cE^V>0$, and $\alpha$ and $\kappa$ satisfy
$\Xi^V-\cE^V> \GG(\alpha/\kappa)$.
Then the stability condition $ \Sigma^V-E^V >0$ holds.
\end{lemma}
In order to prove Lemma \ref{P4.2}, we prepare two lemmas.
We set
\begin{align} \label{sa5}
 E_\mathrm{diag} = \frac{\alpha^2}{2}\sum_{j=1}^N \|\la _j/\sqrt\omega\|^2.
\end{align}
\begin{lemma} \label{enebound1}
 For all $\beta \subset C_N$, it follows that
\begin{align}
E^\#(\beta) & \leq \cE^\#(\beta) + \frac{\alpha^2}{2} \sum_{j\in\beta}
 \|\la _j/\sqrt{\ome} \|^2,\quad \#=0,V.
\end{align}
In particular, it holds that
$\Xi^V\leq \Sigma ^V+E_\mathrm{diag}$.
\end{lemma}
\proof
See Proposition \ref{yui} in Appendix \ref{appb}.
 \qed
\begin{lemma}{\label{enebound2}}
For all $\kappa>0$ it follows that
$ E^V \leq \cE^V + \GG(\alpha/\kappa) - E_\mathrm{diag}$,
\end{lemma}
\proof
For arbitrary $\epsilon>0$, we can choose a normalized vector
$v \in C_0^\infty(\BRN)$ such that
$
 \|(\heff^V - \cE^V) v\| \leq \epsilon$.
Set $\Psi=v\otimes\Omega$.
Then, by Lemma \ref{P4.1}, we have
\begin{align*}
 E^V
 \leq {\mathcal E}^V + \epsilon + \lk \Psi,
 \lk
- E_\mathrm{diag}+\sum_{j=1}^N
 \Delta \Omega_j\rk
\Psi \rk .
\end{align*}
Since $\pi_j$ commutes with $p_i$, $i\neq j$, by setting $T_j=\alpha\pi_j/\kappa$,
we can see that $\Delta\Omega_j = e^{-iT_j}\Omega_j e^{iT_j}-\Omega_j$ and
\begin{align*}
 |\lk \Psi,
 \Delta\Omega_j\Psi \rk |
= |\lk (e^{iT_j}-1)\Psi, \Omega_j e^{iT_j} \Psi \rk + \lk \Psi, \Omega_j(e^{iT_j}-1)\Psi \rk |.
\end{align*}
Hence we have
\begin{align*}
 |\lk \Psi, \Delta\Omega_j \Psi \rk |
\leq
 \frac{|\alpha|}{\kappa} \|\pi_j \Psi \| \cdot \|\Omega_j e^{iT_j}\Psi\|
   + \frac{|\alpha|}{\kappa} \|\pi_j \Psi \| \cdot \|\Omega_j \Psi\|.
\end{align*}
The right-hand side above is identical with
$$\d
 = \frac{|\alpha|}{\sqrt{2}\kappa} \|\la _j/\ome\|
 \left(\lk \Psi, \lk p_j + \frac{|\alpha|}{\kappa}A_j\rk ^2 \Psi \rk ^{1/2}
      + \lk \Psi, p_j ^2 \Psi \rk ^{1/2} \right).
$$
Then we have
\[
 |\lk \Psi, \Delta\Omega_j \Psi \rk |
\leq \frac{|\alpha|}{\sqrt{2}\kappa} \|\la _j/\ome\|
 \left( 2 \| \Omega_j \Psi \| + 2m_j
 + \frac{\sqrt{2}|\alpha|}{\kappa} \| |k|\la _j/\omega \| \right)
 \]
and
\begin{align*}
 E^V&
\leq
\cE^V +\epsilon
 + \sum_{j=1}^N \frac{|\alpha|}{\sqrt{2}\kappa}
\|\la_j/\ome\|
\lk
2m_j+\frac{\sqrt{2}|\alpha|}{\kappa}\| |k|\la _j/\omega\|
\rk
\\
&\quad + \sum_{j=1}^N \frac{\sqrt{2}|\alpha|}{\kappa}
\|\la _j/\ome\|
  \left( c^V (|\cE^V| + \epsilon) + d^V\right)
  -E_\mathrm{diag}.
\end{align*}
Since $\epsilon>0$ is arbitrary, the lemma follows.
\qed
\noindent\textit{Proof of Lemma \ref{P4.2}:} By Lemmas \ref{enebound1} and \ref{enebound2}, we have
\begin{align} \label{saigo}
 \Sigma^V -E^V \geq \Xi^V - \cE^V - \GG(\alpha/\kappa)>0 .
 \end{align}
Then the lemma is proven.
\qed

\subsection{Exponential decays}
It is proven that the functional integration is a strong tool to show an exponential localization of bound state in quantum mechanics.
That can be also applied in quantum field theory.

Let $\pro X= (X_t^1,\dots,X_t^N)_{t\geq 0}$
 be the $N$ independent $d$-dimensional L\'evy processes
 on a probability space $({\cal D}, {\cal B}, \PPPP^x)$, $x\in\BRN$, with
 the characteristic function
\begin{align} \label{char}
\Ebb_\PPPP^0[e^{-iu\cdot X_t}]=
 e^{-t\sum_{j=1}^N (\sqrt{u_j^2+m_j^2}-m_j)},\quad u=(u_1,...,u_N)\in\BRN.
 \end{align}
 Here and in what follows $\Ebb_m^x[\cdots]$ denotes the expectation with respect to a path measure $m^x$ starting from $x$.
Let $\weff=\weff(x_1,..,x_N)=\sum_{j=1}^N V(x_j)+\eff(x)$.
 \begin{proposition} \label{b2}
There exists $\s_0>0$ such that for all $\s\leq \s_0$,
\begin{align} \label{l777}
\|\gr(X)\|_\fff
\leq
e^{t( E^V + E_\mathrm{diag}+\epsilon(\s))}
\lk
\Ebb_\PPPP^X \left[
e^{-2\int_0^t \weff(X_s) ds} \right]
\rk^\han
 \|\gr\|_\hhh
\end{align}
for each $X\in\BRN$, where $\epsilon(\s)>0$ satisfies $\lim_{\s\to0}\epsilon(\s)=0$.
\end{proposition}
\proof See Proposition \ref{ito}.
\qed
From Proposition \ref{b2} it suffices to estimate
$ e^{t( E^V + E_\mathrm{diag})} \Ebb_\PPPP^X \left[ e^{-2\int_0^t \weff(X_s) ds} \right]^\han $
for the exponential decay of $\|\gr(X)\|_\fff$.
To estimate this we divide $\weff$ into two parts. 
Let
\begin{equation*}
B_R=\{x=(x_1,...,x_N)\in \BRN||x|\geq 2R \mbox{ and } \min\{|x_i-x_j|,i\not= j\}\leq |x|/2\}.
\end{equation*}
Define
$V_{\rm eff, \infty}^R=\eff \one_{B_R}$ and
$V_{\rm eff, 0}^R=\eff \one_{B_R^c}$.
Then
\begin{align} \label{kosi6}
\weff = V+V_{\rm eff, 0}^R+V_{\rm eff, \infty}^R.
\end{align}
By the Riemann Lebesgue lemma $\lim_{|x|\to\infty}W_{ij}(x)=0$.
Then notice that
\begin{align*}
&\lim_{|x|\to\infty} (V(x)+V_{\rm eff, 0}^R(x) )=0,\\
 &\|V_{\rm eff, \infty}^R\|_\infty
\leq \frac{\alpha^2}{2}
\sum_{i\not = j} \int \frac{\la_i(k)\la_j(-k)}{\omega(k)}dk.
\end{align*}
The L\'evy measure $\nu_j(dx)=\nu_j(x) dx $ associated with
 the L\'evy process $\pro {X^j}$ is given by
\begin{align} \label{l1}
\nu_j(x)=2\lk\frac{m_j}
{2\pi}\rk^{\frac{d+1}{2}}
\frac{1}{|x|^{\frac{d+1}{2}}}
\int _0^\infty \xi^{\frac{d-1}{2}}
e^{-\half(\xi+\xi^{-1})m_j|x|}
d\xi,\quad x\in\BR.
\end{align}
We note that
$
\nu(x)\leq
C e^{-c |x|}$
with some constants $C, c \geq 0$.
\begin{proposition} \label{l4}
There exist $\eta>0$, $C_1>0$ and $C_2>0$ such that
\begin{align} \label{l5}
\PPPP^0\lk \sup_{0\leq s\leq t}|X_s|>a\rk
\leq
C_1e^{-\eta a}e^{C_2 t}
\end{align}
for all $a>0$.
\end{proposition}
\proof
We see that
\begin{align*}
\PPPP^0\lk \sup_{0\leq s\leq t}|X_s|>a\rk
=
\Ebb_\PPPP^0
\left[
\one_{\sup_{0\leq s\leq t}|X_s|-a>0}
\right]
\leq
e^{-\eta a}
\Ebb_\PPPP^0
\left[e^{\eta \sup_{0\leq s\leq t}|X_s|}
\right].
\end{align*}
It is known that
$\Ebb_\PPPP^0[e^{\eta (
\sup_{0\leq s\leq t}|X_s|)}]<C_1 e^{C_2t}$ for sufficiently small $0<\eta$ \cite{cms90}.
Hence the proposition follows.
\qed
We define
$\BBB=\{X_s\in B_R^c \mbox{ for all } 0\leq s\leq t\}$.
Since $V_{\rm eff, \infty}^R(X_s)=0$ for on $\BBB$,
we have
\begin{align} \label{kosi22}
\Ebb_\PPPP^X\left[e^{-2\int_0^t \weff(X_s) ds}\right]
=
\Ebb_\PPPP^X\left[\one_\BBB e^{-2\int_0^t (V+V_{\rm eff,0}^R)(X_s) ds}\right]+
\Ebb_\PPPP^X\left[\one_{\BBB^c}e^{-2\int_0^t \weff(X_s) ds}\right]
\end{align}
By the Schwartz inequality
\begin{align} \label{kosi}
\Ebb_\PPPP^X \left[ \one_{\BBB^c}e^{-2\int_0^t \weff(X_s) ds}\right]
\leq
\Ebb_\PPPP^X \left[\one_{\BBB^c} e^{-4\int_0^t V_{\rm eff, \infty}^R(X_s) ds}\right]^\han
\Ebb_\PPPP^X \left[\one_{\BBB^c} e^{-4\int_0^t (V+V_{\rm eff,0}^R)(X_s) ds}\right]^\han.
\end{align}
We will estimate terms in \eqref{kosi22} and \eqref{kosi}.
Set
\begin{align*}
& W_a^R(x)=\inf\{
V(y)+V_{\rm eff,\infty}^R(y) ||x-y|<a\},\\
&
W_\infty^R=\inf_{x\in\BRN}
(V(x)+V_{\rm eff,\infty}^R(x)).
\end{align*}
\begin{lemma} \label{l77}
Suppose (1) of (EN). 
Let $R>0$ and $a>0$.
Then for all $X \in\BRN$ and $t>0$ it follows that
\begin{align} \label{l8}
 \Ebb_\PPPP^X
 [e^{-2\int_0^t
(V(X_s)+V_{\rm eff,\infty}^R(X_s) )ds}]
 \leq
 e^{-2t W_a^R(x)}+
C_1 e^{-2tW_\infty^R }
e^{C_2t}e^{-\eta a},
\end{align}
where $C_1,C_2$ and $\eta$ are given in \eqref{l5}.
\end{lemma}
\proof
Set $ A = \{
\sup_{0\leq s\leq t} | X_s| < a \}\subset{\cal D}$.
Since $\pro X$ under the probability measure $\PPPP^X$ and $(X_t+X)_{t\geq0}$ under $\PPPP^0$ are
identically distributed,
we have the identity:
$ \Ebb_\PPPP^X \left[
e^{-2\int_0^t
(V(X_s)+V_{\rm eff,\infty}^R(X_s) ) ds} \right]
=
\Ebb_\PPPP^0\left[
e^{-2\int_0^t
(V(X_s+X)+V_{\rm eff,\infty}^R(X_s+X) ) ds} \right]
$.
Then
we have
\begin{align*}
 \Ebb_\PPPP^0\left[
\one_{A}e^{-2\int_0^t
(V(X_s+X)+V_{\rm eff,\infty}^R(X_s+X) )
 ds} \right]
&\leq e^{-2t W_a^R(x)},\\
\Ebb_\PPPP^0\left[
\one_{A^c}e^{-2\int_0^t
(V(X_s+X)+V_{\rm eff,\infty}^R(X_s+X) ) ds}\right]&
\leq
e^{-2t W_\infty^R}
\Ebb_\PPPP^0\left[
\one_{A^c}
\right]
\leq
e^{-2t W_\infty^R}
C_1e^{C_2 t}e^{-\eta a}
\end{align*}
by Proposition \ref{l4}. Then the lemma follows.
\qed
\begin{lemma} \label{kosi2}
Let $X\in\BRN$ and set $R=|X|$.
Then it follows that
\begin{align} \label{kosi3}
\Ebb_\PPPP^X[\one_{\BBB^c}e^{-4\int_0^t V_{\rm eff, \infty}^R(X_s) ds}]\leq
e^{4\|V_{\rm eff, \infty}\|_\infty t }C_1e^{C_2t}e^{-\eta R},
\end{align}
where $C_1,C_2$ and $\eta$ are given in \eqref{l5}.
\end{lemma}
\proof
Since 
$\Ebb_\PPPP^X[e^{-4\int_0^t V_{\rm eff, \infty}^R(X_s) ds}]
 \leq
 \Ebb_\PPPP^X[e^{4\| V_{\rm eff, \infty}\|_\infty \int_0^t \one_{{B_R}}(X_s) ds}]$,
we can see that
\begin{align*}
&\Ebb_\PPPP^X[e^{-4\int_0^t V_{\rm eff, \infty}^R(X_s) ds}]
\leq
\sum_{n=0}^\infty \frac{(4\| V_{\rm eff, \infty}\|_\infty)^n}{n!}
\int_0^t ds_1\cdots\int_0^t ds_n \Ebb_\PPPP^X\left[
\one_{\BBB^c}
\prod_{j=1}^n \one_{{B_R}}(X_{s_j})\right]\\
&=
\Ebb_\PPPP^X
\left[
\one_{\BBB^c}
\right]
+
\sum_{n=1}^\infty \frac{(4\| V_{\rm eff, \infty}\|_\infty)^n}{n!}
\int_0^t ds_1\cdots\int_0^t ds_n
\Ebb_\PPPP^{{0}}
\left[
\one_{\BBB^c}
\prod_{j=1}^n \one_{B_R}(X+X_{s_j})\right]
\end{align*}
We see that
\begin{align} \label{sala}
\Ebb_\PPPP^X
\left[ \one_{\BBB^c} \right]
\leq \PPPP^0(\sup_{0\leq s\leq t}|X_s+X|>2R)\leq
\PPPP^0 (\sup_{0\leq s\leq t}|X_s|>2R-|X|)=
\PPPP^0 (\sup_{0\leq s\leq t}|X_s|>R).
\end{align}
By the definition of $B_R$ in a similar way to above we have
\begin{align*}
&
\Ebb_\PPPP^X[e^{-4\int_0^t V_{\rm eff, \infty}^R(X_s) ds}]\\
& \leq
\PPPP^X(\BBB^c)+
\sum_{n=1}^\infty \frac{(4\| V_{\rm eff, \infty}\|_\infty)^n}{n!}
\int_0^t\!\! ds_1\!\!\cdots\!\!\int_0^t\!\! ds_n \PPPP^0(|X_{s_1}+X|>2R,\!\!\cdots\!\!,|X_{s_n}+X|>2R)\\
&\leq
\PPPP^X(\BBB^c)+
\sum_{n=1}^\infty \frac{(4\| V_{\rm eff, \infty}\|_\infty)^n}{n!}
\int_0^t ds_1\cdots\int_0^t ds_n \PPPP^0(|X_{s_1}|>R,\cdots,|X_{s_n}|>R).
\end{align*}
By $\PPPP^0(|X_{s_1}|>R,\cdots,|X_{s_n}|>R) \leq 
\PPPP^0\lk \sup_{0\leq s\leq t} |X_{s}|>R\rk$ and
Proposition \ref{l4}, we have
\begin{align*}
&\Ebb_\PPPP^X[e^{-4\int_0^t V_{\rm eff, \infty}^R(X_s) ds}]\\
&\leq
\PPPP^0\lk \sup_{0\leq s\leq t} |X_{s}|>R\rk+\sum_{n=1}^\infty \frac{(4\||V_{\rm eff, \infty}\|_\infty)^n}{n!}
\int_0^t ds_1\cdots\int_0^t ds_n \PPPP^0\lk \sup_{0\leq s\leq t} |X_{s}|>R\rk \\
&\leq
\sum_{n=0}^\infty \frac{(4\||V_{\rm eff, \infty}\|_\infty)^n}{n!}
t^n
C_1 e^{C_2 t} e^{-\eta R}\\
&=
e^{4\||V_{\rm eff, \infty}\|_\infty t}
C_1 e^{C_2 t} e^{-\eta R}.
\end{align*}
Hence the lemma follows.
\qed
\begin{lemma} \label{expdecay}
Let $\gr$ be the infrared regularized
ground state.
Suppose (1) of (EN)
and
$E^V+E_\mathrm{diag} <0$.
Furthermore we assume that
 $E^V+E_\mathrm{diag}+\epsilon(\s)<-\gamma$ with some $\gamma>0$ for $\s<{\bar\s}$,
 where $\epsilon(\s)$ is given in Proposition \ref{b2}.
Then there exist $\delta>0$ and $C_\delta>$ independent of $\s$ such that
\begin{align} \label{l9}
\sup_{0<\s<{\bar\s}}
\|\gr(X)\|_\fff\leq
{C_\delta}e^{-\delta \min \{\gamma,\eta\}|X|},
\end{align}
where $\eta>0$ is given in Proposition \ref{l4}.
\end{lemma}
\proof
We set $\tilde E=E^V+E_\mathrm{diag}+\epsilon(\s)$.
It is enough to estimate
$e^{2t\tilde E} \Ebb_\PPPP^X \left [e^{-2\int_0^t \weff(X_s) ds}\right]$
by Proposition \ref{b2}.
Recall that
 $W_a^R(x)=\inf\{W^R(y)||x-y|\leq a\}$.
Then
\begin{align} \label{ln1}
\lim_{|x|\to\infty}W_{|x|/2}^{|x|}(x)=0.\end{align}
Hence there exists a positive constant $R^\ast$ such that
$|W_{|X|/2}^{|X|}(X) |\leq |\tilde E|/2$ for all $X$ such that $|X|>R^\ast$.
Suppose that $|X|>R^\ast$ and let
 $R=|X|$. We divide $\weff$ as in \eqref{kosi6} for $R$.
We have
\begin{align*}
&e^{2t\tilde E}\Ebb_\PPPP^X \left[
e^{-\int_0^t \weff(X_s) ds}
\right]\\
&\leq
e^{2t\tilde E}
\Ebb_\PPPP^X \left[
\one_{\BBB}
e^{-2\int_0^t (V+V_{\rm eff,0}^R)(X_s) ds}
\right]\\
&+
e^{2t\tilde E}
\lk
\Ebb_\PPPP^X \left[
\one_{\BBB^c}
e^{-4\int_0^t (V+V_{\rm eff,0}^R)(X_s)ds}
\right]\rk
^\han
\lk
\Ebb_\PPPP^X \left[\one_{\BBB^c}
e^{-4\int_0^t (V+V_{\rm eff,\infty}^R)(X_s) ds}
\right]\rk^\han
\end{align*}
Two terms
$\Ebb_\PPPP^X \left[\one_\BBB e^{-2\int_0^t (V+V_{\rm eff,0}^R)(X_s) ds}\right]$
 and
$\Ebb_\PPPP^X \left[\one_{\BBB^c} e^{-4\int_0^t (V+V_{\rm eff,0}^R)(X_s) ds}\right]$
 can be estimated as
\begin{align}
&\Ebb_\PPPP^X\left[\one_\BBB e^{-2\int_0^t (V+V_{\rm eff,0}^R)(X_s) ds}\right]
\leq
 e^{-2t W_a^R(x)}+
C_1 e^{-2tW_\infty^R }
e^{C_2t}e^{-\eta a},
\\
&
\Ebb_\PPPP^x\left[\one_{\BBB^c} e^{-4\int_0^t (V+V_{\rm eff,0}^R)(X_s) ds}\right]
\leq
 e^{-4t W_a^R(x)}+
C_1 e^{-4tW_\infty^R }
e^{C_2t}e^{-\eta a}
\end{align}
by Lemma \ref{l77}.
Let us set $t=t(X)=\eps|X|$ and $a=|X|/2$.
Then we can see that
$W_{|X|/2}^{|X|}(X) -\tilde E>-\tilde E/2>0$,
since $\tilde E<0$ by assumption.
Hence
\begin{align*}
e^{2t\tilde E}
\Ebb_\PPPP^X\left[\one_\BBB e^{-2\int_0^t (V+V_{\rm eff,0}^R)(X_s) ds}\right]
&\leq
e^{\eps \tilde E|X|}+C_2e^{\eps C_2|X|-\eta |X|/2-2\eps W_\infty^{|X|}|X|}\\
&\leq
e^{-\eps \gamma |X|}+C_2
e^{-(\eta/2 +2\eps W_\infty^{|X|} -\eps C_2)|X|}.
\end{align*}
Similarly we have
\begin{align*}
&e^{4t\tilde E}
\Ebb_\PPPP^X\left[\one_\BBB e^{-4\int_0^t (V+V_{\rm eff,0}^R)(X_s) ds}\right]
\leq
e^{-2\eps \gamma |X|}+C_2
e^{-(\eta/2 +4\eps W_\infty^{|X|} -\eps C_2)|X|}.
\end{align*}
Finally by Lemma \ref{kosi2} we have
\begin{align*}
e^{4t\tilde E}
\Ebb_\PPPP^X\left[\one_\BBB e^{-4\int_0^t \weff(X_s) ds}\right]
&\leq
C_1e^{4\eps \tilde E +4\|V_{\rm eff,\infty}\|_\infty \eps +C_2\eps-\eta)|X|}\\
&\leq
C_1e^{-(4\eps \gamma -4 \|V_{\rm eff,\infty}\|_\infty \eps - C_2\eps+\eta)|X|}.
\end{align*}
Note that $W_\infty^{|X|}\to 0$ as $|X|\to\infty$.
Take sufficiently small $\eps>0$ such that
$\eta/2+(2W_\infty^{|X|}-C_2)\eps>0$, $\eta/2+(4W_\infty^{|X|}-C_2)\eps>0$ and
$(4\gamma -4 \|V_{\rm eff,\infty}\|_\infty - C_2)\eps+\eta>0$, then
$\|\gr(X)\|_\fff\leq D_1e^{-\min\{\eta,\gamma\}D_2|X|}$ follows.
 Then the lemma is proven.
 \qed
\begin{corollary} \label{b3}
Suppose (1) of (EN). Then \eqref{l9} holds for sufficiently small $|\alpha/\kappa|$.
\end{corollary}
\proof
Notice that
 $E^V\leq {\cal E}^V+\GG(\alpha/\kappa) -E_\mathrm{diag}$ in Lemma \ref{enebound2}.
Since ${\cal E}^V<0$ and $\lim_{t\to 0}\GG(t)=0$, the corollary follows.
\qed

\subsection{Proof of Theorem \ref{main}\   and  Theorem \ref{enhanced}}

\subsubsection{Proof of Theorem \ref{main}}

{\it Proof of Theorem \ref{main}}:

Note that $0<\cE^V-\Xi^V$ is equivalent to $\inf\sigma(H^V_{\rm eff})\in \sigma_{\rm disc}(H^V_{\rm eff})$.
Uniform exponential decay $\|\gr(x)\|_\fff \leq C_\delta e^{-\delta|x|}$
 is shown for sufficiently small $|\alpha/\kappa|$ in Lemma \ref{expdecay}.
Then by 
$
 \Sigma^V -E^V \geq \Xi^V - \cE^V - \GG(\alpha/\kappa)
 $
  and  the fact that $\lim_{\kappa\to\infty }\GG(\alpha/\kappa )=0$,
there exists $\kappa_0$ such that 
for arbitrary $\kappa>\kappa_0$ 
the stability condition $E^V<\Sigma^V$ holds.
Therefore, by Proposition \ref{P3.1}, $H^V$ has a ground state.
\qed
\subsubsection{Proof of Theorem \ref{enhanced}}
Now we show the enhanced binding.
It is enough to show $ \cE^V < \Xi^V$, since the uniform exponential 
decay $\|\gr(x)\|_\fff<C_\delta e^{-\delta|x|}$ is established by 
Proposition \ref{P3.1}.
\begin{lemma} \label{ene}
 Let $\beta\subsetneqq \N $ but $\beta\neq\emptyset$.
Then there exists $\alpha_1>0$ such that,
for all $\alpha$ with $|\alpha|>\alpha_1$,
$ \cE^0 < \cE^V(\beta)+ \cE^0(\beta^c)$.
In particular $\cE^0<\Xi^V$ holds
for $|\alpha|>\alpha_1$.
\end{lemma}
\proof
We have
\begin{align*}
 & \cE^0 = \alpha^2 \sum_{i<j} W_{ij}(0) + o(\alpha^2), \quad
 \cE^V(\beta) = \alpha^2 \sum_{\atopb{i<j}{i,j\in\beta}}W_{ij}(0) + o(\alpha^2), \\
 & \cE^0(\beta^c) = \alpha^2 \sum_{\atopb{i<j}{i,j\in\beta^c}}W_{ij}(0) + o(\alpha^2) .
\end{align*}
Since
$\d \sum_{\atopb{i<j}{i\in\beta, j\in\beta^c}}W_{ij}(0)+
 \sum_{\atopb{i<j}{i\in\beta^c, j\in\beta}}W_{ij}(0)<0$,
the lemma holds.
\qed
To see the enhanced binding we want to investigate the center of motion of $\heff^V$. Notice
that $\heff^0$ commutes with the total momentum $\pt=\sum_{j=1}^N p_j$. Then it can be decomposable with respect to the spectrum of $\pt$.
Let
$ \UU = e^{ix_1\cdot \sum_{j=2}^N p_j }$, which
 diagonalize $\pt$ as
 $\UU \pt \UU^{-1} =p_1$. Hence it also diagonalize $\heff^0$, and we obtain
 that
 \begin{align*}
& \UU \heff^0 \UU ^{-1} = \Omega_1\lk
p_1-\sum_{j=2}^Np_j
\rk + \sum_{j=2}^N \Omega_j(p_j ) + \sum_{j\geq 2} \alpha^2W_{1j}(x_j) +\!\! \sum_{2\leq i<j\leq N} \alpha^2 W_{ij}(x_i-x_j),\\
& \UU \heff^V\UU ^{-1} = \heff^0 +V(x_1) + \sum_{j=2}^N V(x_1+x_j).
\end{align*}
Then we have
\begin{align*}
& \UU \heff^0\UU ^{-1} = \int_{\BR}^\oplus k(P) dP, \\
& k(P) = \Omega_1\lk
P-\sum_{j=2}^Np_j
\rk
 + \sum_{j=2}^N \Omega_j(p_j )
  + \sum_{j\geq 2} \alpha^2 W_{1j}(x_j) + \sum_{2\leq i<j\leq N} \alpha^2 W_{ij}(x_i-x_j).
\end{align*}

\begin{lemma} \label{p=0}
It follows that ${\cal E}^0=\inf\!\sigma(k(0))$.
\end{lemma}
\proof
Set $\inf\!\sigma(k(P))=E(P)$ for simplicity.
It can be seen
in Appendix \ref{appc} that
\begin{align} \label{ebound}
E(0)\leq E(P)
\end{align}
 holds
for all $P$, and that $E(P)$ is continuous in $P$.
Then it follows that
$\d (\Phi, H\Phi)=\int_\BR (\Phi(P), k(P)\Phi(P))dP\geq E(0)\|\Phi\|^2$ for $\Phi\in D(H)$. Then $E(0)\leq {\cal E}^0$. On the other hand
let us set $\Phi_\epsilon =\int^\oplus_\BR \Phi(P)
\one_{[0,\epsilon)}(P)dP$.
 We have
$\|\Phi_\epsilon\|^2
{\cal E}^0\leq (\Phi_\epsilon, H\Phi_\epsilon)\leq \sup_{|P|<\epsilon}E(P)\|\Phi_\epsilon\|^2$. Take $\epsilon\downarrow 0$ on both sides we have ${\cal E}^0\leq E(0)+\delta$ for arbitrary $\delta>0$, since $E(P)$ is continuous in $P$.
Hence $E(0)\geq {\cal E}^0$ and then ${\cal E}^0=E(0)$ follows.
\qed
\begin{lemma} \label{gr}
There exists $\alpha_2(P)>0$ such that
$\inf\!\sigma(k(P))\in \s_{\rm disc}(k(P))$ for every $P\in\BR$ for $|\alpha|>\alpha_2(P)$.
 In particular $k(0)$ has a ground state for
 $|\alpha|>\alpha_2$ with some $\alpha_2>0$.
\end{lemma}
\proof
Notice that $W_{ij}(0)<0$, $W_{ij}(x) > W_{ij}(0)$ for $x\not =0$, and $\lim_{|x|\to\infty}W_{ij}(x)=0$.
Set $\mathbf{X}=(x_2,\dots,x_N)$.
Let $a=\{2,...,N\}$.
Let $\{\til{j}_\beta\}_{\beta \subset a }$ be the Ruelle-Simon partition of unity \cite[Definition 3.4]{cfks87},
i.e., $\til j_\beta(\lambda \mathbf{X})=\til j_\beta(\mathbf{X})$ for all $\lambda>1$, $|\mathbf{X}|=1$, and there exists a constant
$C>0$ such that
${\rm supp}\til j_\beta\cap \{\mathbf{X}||\mathbf{X}|>1\}\subset\{\mathbf{X}||\mathbf{X}_i-\mathbf{X}_j|\geq C|\mathbf{X}|\mbox{for all }(ij)\not\subset \beta\}$.
 We set $j_\beta(\mathbf{X})= \til{j}_\beta(\mathbf{X}/R)$.
 Then
\begin{align}
 k(P) = j_{a}k(P)j_{a} +
   \sum_{\beta \subsetneq a} j_\beta k(P) j_\beta
   + o(\one), \label{imsk}
\end{align}
where $o(\one)$ denotes a bounded operator such that $\lim_{R\to\infty} \norm{o(\one)}=0$.
We set
\begin{align*}
 &k_\beta = \sum_{j\in\beta}(\Omega_j(p_j)+\alpha^2 W_{1j}(x_j)) + \sum_{i,j\in\beta} \alpha^2 W_{ij}(x_i-x_j)\\
 &\bar{k}_{\beta^c} = \sum_{j\in\beta^c} \Omega_j(p_j) + \sum_{i,j\in\beta^c} \alpha^2 W_{ij}(x_i-x_j)
\end{align*}
With the identification $L^2(\RR^{d(N-1)}) \cong L^2(\RR^{d|\beta|})\tensor L^2(\RR^{d|\beta|^c)})$,
we can write
\begin{align}
 j_\beta k(P) j_\beta
 = j_\beta \Omega_1\lk P-\sum_{j=2}^Np_j\rk j_\beta
  + j_\beta ( k_\beta \tensor \one + \one \tensor \bar k_{\beta^c}) j_\beta + I_\beta j_\beta^2 \label{imsk2}
\end{align}
where
$ I_\beta = \sum_{j\in\beta^c}\alpha^2 W_{1j}(x_j) + \sum_{\atopb{i\in \beta, j\in\beta^c}{i\in\beta^c,j\in\beta}}
     \alpha^2 W_{ij}(x_i-x_j)$.
Hence, \eqref{imsk} and \eqref{imsk2} imply
\begin{align*}
 k(P) \geq E_0(k(P))j_{a}^2
   + \sum_{\beta \subsetneq a} j_\beta(k_\beta \tensor \one + \one \tensor \bar{k}_{\beta^c}
    + I_\beta) j_\beta + o(\one).
\end{align*}
Note that $j_{a}^2$ and $I_\beta j_\beta^2$ are relatively compact with respect to $k(P)$.
Thus we have
\begin{align*}
 \inf \sigma_\mathrm{ess}(k(P)) \geq \max\{E_0(k_\beta) + E_0(\bar{k}_{\beta^c}) |
   \beta \subsetneq a\}.
\end{align*}
For all $\beta\subsetneq a$ it holds that
\begin{align}
& \lim_{\alpha\to\infty} \frac{E_0(k(P))}{\alpha^2} = \sum_{i<j}W_{ij}(0)
 < \sum_{j\in \beta} W_{1j}(0) + \sum_{\atopb{i,j \in\beta }{i<j}}W_{ij}(0)
    + \sum_{\atopb{i,j\in \beta^c}{i<j}}W_{ij}(0) \\
& = \lim_{\alpha\to\infty} \frac{E_0(k_\beta)+E_0(\bar{k}_{\beta^c} ) }{\alpha^2}.\non
\end{align}
Therefore there exist $\alpha_2(P)$ such that for all $\alpha>\alpha_2(P)$,
$ \inf\sigma_\mathrm{eff}(k(P)) > E_0(k(P))$.
 \qed

\begin{lemma}{\label{delta}}
Let $|\alpha|>\alpha_2$, where $\alpha_2$ is given in Lemma \ref{gr}, and
 $u_\alpha$ be a normalized ground state of $k(0)$.
Then
$ |u_\alpha(x_2,\ldots,x_N)|^2 \to \delta(x_2)\cdots\delta(x_N)$ as $\alpha\to\infty$ in the
sense of distributions.
\end{lemma}
\proof
It suffices to show that for all $\epsilon>0$,
\begin{align}
 \lim_{\alpha\to\infty} \int_{|\BX |>\epsilon} |u_\alpha(\BX )|^2 d\BX =0, 
 \label{2.2400}
\end{align}
where $\BX =(x_2,\cdots,x_N)$,
since \eqref{2.2400} implies that
$$\d\lim_{\alpha\to0}
 \int_{\RR^{d(N-1)}}f(\BX )|u_\alpha(\BX )|^2 d\BX
 = f(0)$$ for
all $f\in C_0^\infty(\RR^{d(N-1)})$.
We write $k_\alpha(0)$ to emphasize the $\alpha$ dependence of $k(0)$.
Since $k_\alpha(0)/\alpha^2 \geq \sum_{i<j}W_{ij}(0)$ and
$\lim_{\alpha\to\infty} \inf\sigma(k_\alpha(0)) /\alpha^2 =\sum_{i<j}W_{ij}(0)$,
we have
\begin{align*}
 \sum_{i<j}W_{ij}(0) =& \lim_{\alpha\to0} \alpha^{-2}(u_\alpha, k_\alpha(0)u_\alpha) \notag \\
 \geq& \liminf_{\alpha\to\infty} \Big(u_\alpha, \Big(\sum_{j\geq2} W_{1j}(x_j) + \sum_{2\leq i<j\leq N}W_{ij}(x_i-x_j)\Big)u_\alpha\Big)
 \geq \sum_{i<j}W_{ij}(0). \label{2.25-3}
\end{align*}
Then
\begin{equation} \label{sasa10}
 \sum_{i<j}W_{ij}(0) =
 \liminf_{\alpha\to\infty} 
 \Big(u_\alpha, \Big(\sum_{j\geq2} W_{1j}(x_j) + \sum_{2\leq i<j\leq N}W_{ij}(x_i-x_j)
 \Big)u_\alpha\Big)
 \end{equation}
follows. Suppose that
$\d c_\epsilon=\liminf_{\alpha\to\infty} \int_{|\BX |>\epsilon} |u_\alpha(\BX )|^2 \d\BX >0$.
Then
\begin{align*}
\liminf_{\alpha\to\infty} \int_{\RR^{d(N-1)}} \! \sum_{j\geq 2}(W_{1j}(x_j)-W_{1j}(0))
 |u_\alpha(\BX )|^2 d\BX
 > c_\epsilon \sum_{j\geq 2}\sup_{|\BX |>\epsilon} (W_{1j}(x_j)-W_{1j}(0)) >0,
\end{align*}
which contradicts \eqref{sasa10}. Therefore \eqref{2.2400} holds.
 \qed

\textit{Proof of Theorem \ref{enhanced}:}

First we assume that $V \in C_0^\infty(\BR)$.
It is enough to show
$\cE^V<\Xi^V$, since the uniform exponential decay $\|\gr(x)\|\leq C_\delta e^{-\delta|x|}$ is established in Lemma \ref{expdecay} for sufficiently small $|\alpha/\kappa|$.
Assume $|\alpha| >\max\{\alpha_1,\alpha_2\}>0$.
Let $u_\alpha$ be a normalized ground state of $k(0)$.
By
$\Omega_1(a+b)\leq |a| + \Omega_1(b)$
 for $a,b\in\BR$,
we have
\begin{align}
 \UU \heff^0\UU ^{-1} \leq \sqrt{-\Delta_1} + k(0). \label{dec5.2}
\end{align}
By (2) of (EN), there exists a normalized vector $v\in C_0^\infty(\BR)$ such that
\begin{align}
( v, (\sqrt{-\Delta} + NV) v ) < 0. \label{yy3}
\end{align}
We set
$\Psi(x_1,\cdots,x_N) = v(x_1)u_\alpha(x_2,\cdots,x_N)$.
Then, by \eqref{dec5.2}
\begin{align}
\label{yy4} \cE^V & \leq (\Psi, \UU \heff^V\UU ^{-1}\Psi)
 \leq ( v, (\sqrt{-\Delta}+V) v) + {\cal E}^0
 + (\Psi,\sum_{j=2}^N V(x_1+x_j) \Psi).
\end{align}
Let
$\d
 V_{j,\mathrm{smeared}}^\alpha(x_1) =
 \int_{\mathbb{R}^{d(N-1)}}
 V(x_j+x_1) |u_\alpha(\BX )|^2d\BX
 $.
By Lemma~\ref{delta}, we have
$$
 \lim_{\alpha\to\infty}( \Psi, V(x_j+x_1) \Psi)
 =\lim_{\alpha\to\infty} ( v, V_{j,\mathrm{smeared}}^\alpha v)
 = ( v, Vv )
$$
 and then
by \eqref{yy3} and \eqref{yy4},
\begin{align}
\label{yyy4} \cE^V
 \leq ( v, (\sqrt{-\Delta}+NV) v) +
 {\cal E}^0<{\cal E}^0
 \end{align}
follows for $\alpha>\alpha_3$ with some $\alpha_3>0$.
By this inequality and Lemma \ref{ene}, we conclude that for $\alpha$ with
$|\alpha|>\bar \alpha=\max\{\alpha_1,\alpha_2,\alpha_3\}$,
\begin{align*}
\Sigma^V-E^V&\geq \Xi^V - {\cal E}^V -\GG(\alpha/\kappa) \geq 
{\cal E}^0 - {\cal E}^V -\GG(\alpha/\kappa)\\
&> -( v, (\sqrt{-\Delta}+NV) v)-\GG(\alpha/\kappa).
\end{align*}
Notice that 
$\GG(\alpha/\kappa)\to 0$ as $\kappa\to\infty$ and 
$-( v, (\sqrt{-\Delta}+NV) v)>0$. Then the right-hand side above is positive for 
sufficiently small $|\alpha|/\kappa$.  
Since $\GG$ is monotonously increasing, it is trivial to see that 
$\kappa_0=\bar\alpha/ \GG^{-1}(a)$, where $a=
-( v, (\sqrt{-\Delta}+NV) v)$ and $\bar\alpha(\kappa)=
\GG^{-1}(a)\kappa$.
Then 
the theorem follows for $V\in C_0^\infty(\BR)$. 
For general $V$ we can prove the theorem by the same limiting argument as
\cite[Appendix]{hs08}. See Appendix \ref{ess}
\qed

\appendix

\section{Stability condition:relativistic version}
\label{appa}
In this section we shall prove Proposition \ref{P3.1}.
We only show an outline of the proof. The detail is left to the reader.
\subsection{Case $\s>0$}
\label{massive}
Since the scaling parameter $\kappa$ does not play any role in this section we set $\kappa =1$.
Let $\sigma>0$.
We decompose the single boson Hilbert space into high energy part and low energy part as
$ L^2(\BR) \cong \cK_{>\sigma} \oplus \cK_{\leq \sigma}$,
where $\cK_{\leq \sigma}=L^2(\{k\in\BR | \ome(k)\leq \sigma\})$ and
$\cK_{>\sigma} = L^2(\{k\in\BR | \ome(k)>\sigma\}).$
Correspondingly, we have the identification:
\begin{align}
 \hhh \cong \hhh _{>\sigma} \tensor \fff(\cK_{\leq \sigma}), \label{id1}
\end{align}
where $\hhh _{>\sigma}=L^2(\BRN)\tensor \fff(\cK_{>\sigma}) $.
We define the regularized Hamiltonian by
\begin{align}
 H^V_\sigma = H_0 + \his .
\end{align}
 Here $\his $ is regularized interaction defined by
$ \his = \sum_{j=1}^N\alpha_j \int^\oplus_{\BRN} \phi_{j,\sigma}(x_j)dX$,
and
 $\phi_{j,\sigma}(x)$ is given by $\phi_j(x)$ with cutoff $\lambda_j(k)$ replaced by
$\lambda_j(k)\one_{\ome(k)>\sigma}(k)$.
Then $H^V_\sigma$ approximates $H^V$ in the following sense:
\begin{lemma}{\label{L3.2}}
 $H^V_\sigma$ converges to $H^V$ as $\sigma\to 0$ in the norm resolvent sense.
\end{lemma}
Let $ E^V_\sigma = \inf\sigma(H^V_\sigma )$ and $\Sigma^V_\sigma$ be 
a lowest two cluster threshold for $H^V_\sigma$, which is defined 
in the same way as $\Sigma^V$.
From Lemma \ref{L3.2}, we can show that $E^V_\sigma$ and $\Sigma^V_\sigma$
converges to $E^V$ as $\Sigma^V$ as $\sigma \to 0$, respectively.
Therefore
for sufficiently small $\sigma>0$, it follows that
\begin{align}
 \Sigma^V_\sigma > E^V_\sigma. \label{bs}
\end{align}
Under the identification \eqref{id1}, $H^V_\sigma$ can be decomposed as
\begin{align*}
 H^V_\sigma \cong H^V_\sigma\lceil _{\hhh _{>\sigma}}
 \tensor \one_{\fff(\cK_{\leq \sigma})}
         +
 \one_{\hhh _{>\sigma}}\tensor
 \hf\lceil_{\fff(\cK_{\leq \sigma})}
\end{align*}
Since $\hf\lceil_{\fff(\cK_{\leq \sigma})}$ has a ground state,
$H^V_\sigma$ also may have a ground state if and only if
$H^V_\sigma\lceil _{\hhh _{>\sigma}}$ does.
We shall prove the existence of a ground state of $H^V_\sigma\lceil _{\hhh _{>\sigma}}$
for sufficiently small $\sigma>0$ in what follows.
For $\sigma > 0$, we truncate $\omega$ as
\begin{align*}
 \omega_{\sigma}(k) = \begin{cases}
			 |k|  \quad &\text{for}\quad  |k|>\sigma \\
			 \sigma \quad &\text{for}\quad |k|\leq \sigma,
		    \end{cases}
\end{align*}
and we set $H_{\mathrm{f},\sigma}=\dG(\omega_\sigma)$.
Then
\begin{align*}
H^V_\sigma\lceil _{\hhh _{>\sigma}}
 = H_{0,\sigma} + \his
\end{align*}
with $ H_{0,\sigma} = H_\mathrm{p}\otimes\one + \one\otimes H_{\mathrm{f},\sigma} $.
We denote the Fourier transformation from
$L^2(\RR^d_y)$ to $L^2(\RR^d_k)$ by $F$.
We set
$ \check\cK_{>\sigma} = \{ \check{f}=F^{-1}f \in L^2(\RR^d_y)| f \in \cK_{>\sigma}\}$.
We introduce a notation.
Let $T :\ck_1\to \ck_2$ be a contraction operator from
a Hilbert space $\ck_1$ to another
one $\ck_2$. Then we define
$\d \Gamma(T) = \oplus_{n=0}^\infty \otimes^n T $ with $\otimes^0 T=\one$, 
which is also a contraction operator from $\fff(\ck_1)$ to $\fff(\ck_2)$.
Let
\begin{align*}
 \check{H}^V_\sigma = \Gamma(F^{-1}) H^V_\sigma\lceil_{\hhh_{>\sigma}} \Gamma(F),
\end{align*}
which is defined on 
$\check{\hhh }_{>\sigma} = L^2 (\RR^{dN})\tensor \fff(\check{\cK}_{>\sigma})$.
Let $\cut , \bar\cut \in C^\infty(\RR^{dN})$ be
a cutoff function such that $\cut (X)^2+\bar\cut (X)^2=1$ with $\cut (X)=1$ 
for $|X|\leq 1$ and $\cut (X)=0$ for $|X|\geq 2$.
Then the following statement holds:
For $R>0$, we set $\cut _R(X)=\cut (X/R)$, $\bar\cut _R(X)=\bar\cut (X/R)$.
\begin{lemma}{\label{L-loc1}}
It follows that
\begin{align*}
 & \check{H}^V_\sigma
 = \cut _R\check{H}^V_\sigma \cut _R + \bar\cut _R \check{H}^V_\sigma \bar\cut _R
   + \hat{O}(R^{-1}),
\end{align*}
 where $\hat{O}(R^{-1})$ is an operator such that $\| \hat{O}(R^{-1})\| \leq C/R$ for some constant $C>0$.
\end{lemma}
\proof
The operator equality
\begin{align} \label{sa14}
 \check{H}^V_\sigma = \cut _R\check{H}^V_\sigma \cut _R + \bar\cut _R \check{H}^V_\sigma \bar\cut _R
 + \frac{1}{2}\sum_{j=1}^N[\cut _R,[\cut _R,\Omega_j(p_j )]]
 + \frac{1}{2} \sum_{j=1}^N[\bar\cut _R,[\bar\cut _R,\Omega_j(p_j )]].
\end{align}
holds. By the Fourier transformation, we have
\begin{align*}
 [\cut _R,\Omega_j(p_j )] =
 (2\pi)^{-dN/2}  \int_{\RR^{dN}} \hat\cut (K) e^{iK\cdot X/R}
 \big(\Omega_j(p_j )-\Omega_j(p_j -k_j/R) \big) dK,
\end{align*}
where $K=(k_1,\cdots,k_N)\in\RR^{dN}$. By the triangle inequality, we have
\begin{align*}
 |\Omega_j(p_j )-\Omega_j(p_j -\frac{k_j}{R})|
 =& \big| \|(p_j ,m_j)\|_{\CC^4} - \|(p_j -\frac{k_j}{R},m_j)\|_{\CC^4} \| \big|
 \leq \| (\frac{k_j}{R},0)\|_{\CC^4} = \frac{1}{R}
 |k_j|.
\end{align*}
Hence, $[\cut _R,\Omega_j(p_j )]$ is a bounded operator with
the bound
\begin{align}
 \| [\cut _R,\Omega_j(p_j )] \|
 \leq \frac{1}{R}
 (2\pi)^{-dN/2}  \int_{\RR^{dN}} |\hat\cut (K)| \cdot |k_j| dK . \label{ims1}
\end{align}
Similarly, by noting that $\one-\bar\cut \in C_0^\infty(\RR^{dN})$ and
$[\bar\cut _R,\Omega_j(p_j )]=[\one-\bar\cut _R,\Omega_j(p_j )]$, we have
\begin{align*}
 \|[\bar\cut _R,\Omega_j(p_j )] \|
 \leq \frac{1}{R} (2\pi)^{-dN/2}  \int_{\RR^{dN}} |\widehat{(\one-\bar\cut (K))}| \cdot |k_j| dK .
\end{align*}
Then the lemma follows.
\qed

Let $j, \bar{j} \in C_0^\infty(\BR)$ be another cutoff function such that
$j(y)^2+\bar{j}(y)^2=1$ for every $y\in\BR$
with $j(y)=1$ for $|y|\leq 1$ and $j(y)=0$ for $|y|\geq 2$.
We set $j_P(y)=j(y/P)$, $\bar{j}_P(y)=\bar{j}(y/P)$ for $P>0$.
The map
\begin{align*}
 u_P : \check\cK_{>\sigma} \to L^2(\RR^d_y)\oplus L^2(\RR^d_y), \quad f \mapsto j_P f\oplus \bar{j}_P f
\end{align*}
is isometry since $\|j_Pf\oplus \bar{j}_P f\|^2 = \|f\|^2$.
We note that $u_P^*$ maps $f\oplus g \in L^2(\BR_y)\oplus L^2(\BR_y) $ to
$j_Pf + \bar{j}_Pg \in L^2(\BR) $.
 The operator
 $$U_P=\one_{\LRT}\otimes \Gamma(u_P):
\check{\hhh }_{>\sigma}\to
\check\hhh \tensor \fff(L^2(\RR^d_y))$$
is also an isometry, where
$\check\hhh = L^2(\RR^{dN})\tensor \fff(L^2(\RR^d_y))$.
Let
 $\check{H}_{0,\sigma}= \Gamma(F^{-1}) H_{0,\sigma}\Gamma(F)$ and
 $\check{H}_{\rm f,\sigma}= \Gamma(F^{-1})\hfs \Gamma(F)$.
 \begin{lemma} {\label{L-loc2}}
For every $\sigma>0$, we have
\begin{align*}
  \cut _R \check{H}^V_\sigma\cut _R
  = \cut _R U_P^* \{ \check{H}^V_\sigma \otimes\one +
  \one\otimes \check{H}_{\mathrm{f},\sigma} \} U_P \cut _R + \hat{o}(\one),
\end{align*}
as operators in $\hhh _{>\sigma}$, where $\hat{o}(\one)$ denotes an operator 
such that $\hat{o}(\one)(\check{H}_{0,\sigma}+\one)^{-1}$ is bounded and
$ \lim_{P\to\infty}\lim_{R\to\infty} \| \hat{o}(\one)(\check{H}_{0,\sigma}+1)^{-1}\| = 0$.
\end{lemma}
\proof
See \cite[Lemma A.1]{gll01}.
\qed
\begin{lemma}{\label{L-sigma}}
We have
$
 \bar\cut _R \check{H}^V_\sigma \bar\cut _R
 \geq \Sigma^V_\sigma\bar\cut _R^2 +o(R^0)$,
where $o(R^0)$ is a number such that $\lim_{R\to \infty}o(R^0)=0$.
\end{lemma}
\proof
See \cite[Lemma A.1]{gll01}.
\qed
\begin{proposition} \label{sasaki1}
There exists a ground state of
 $H_\sigma^V$.
\end{proposition}
\proof
By Lemma \ref{L-loc1} and Lemma \ref{L-loc2},
\begin{align*}
 \check{H}^V_\sigma
 = \cut _R U_P^*\{\check{H}^V_\sigma \otimes\one + \one\otimes \check{H}_{\mathrm{f},\sigma}\} U_P \cut _R
   + \bar\cut _R \check{H}^V_\sigma\bar\cut _R + \hat{o}(\one).
\end{align*}
Since $\omega_\sigma\geq \sigma$, we have $\check{H}_{\mathrm{f},\sigma}\geq \sigma(\one-P_\Omega)$,
where $P_\Omega$ denotes the orthogonal projection on the vacuum space $\{\mathbb{C}\Omega\}$.
By this inequality and Lemma \ref{L-sigma},
\begin{align*}
 \check{H}^V_\sigma
 \geq
 (E^V_\sigma+\sigma) \cut _R^2 + \Sigma^V_\sigma \bar\cut _R^2
    -K     + \hat{o}(\one),
\end{align*}
where
 $K=\sigma \cut _RU^{-1}_P (\one\otimes P_\Omega)U_P\cut _R=\cut _R^2\tensor\Gamma(j_P)$.
$K$ is relatively compact with respect to $\sum_{j=1}^N\Omega_j +\check{H}_{\mathrm{f},\sigma}$.
Since, by (V), $\sum_{j=1}^N\Omega_j+\check{H}_{\mathrm{f},\sigma}$ is also relatively bounded
with respect to $\check{H}^V_\sigma$,
$K$ is then relatively compact with respect to $\check{H}^V_\sigma$.
By the definition of $\hat{o}(\one)$,
there is a constant $C$ independent of $P$ and $R$
such that
$ \hat{o}(\one) \geq -o(\one) (\check{H}^V_\sigma +C)$.
Thus, we have the operator inequality
\begin{align*}
(1+o(\one)) \check{H}^V_\sigma -E^V_\sigma + o(\one) - K
 \geq \sigma \cut _R^2 + (\Sigma^V_\sigma -E^V_\sigma)\bar\cut _R^2
  \geq \min\{\sigma, \Sigma^V_\sigma - E^V_\sigma\}.
\end{align*}
Since $K$ does not change the essential spectrum of $\check{H}^V_\sigma$,
 for all $P$ and $R$, we have
 \begin{align*}
 (1+o(\one)) \inf(\sigma_\mathrm{ess}(H^V_\sigma)) -E^V_\sigma +o(\one) \geq \min\{\sigma, \Sigma^V_\sigma - E^V_\sigma\}.
 \end{align*}
Hence, by \eqref{bs},
\begin{align*}
  \inf \sigma_\mathrm{ess}(H^V_\sigma) - E^V_\sigma \geq \min\{\sigma, \Sigma^V_\sigma - E^V_\sigma\} >0.
\end{align*}
Therefore $\sigma(\check{H}^V_\sigma)\cap [E^V_\sigma, E^V_\sigma+\min\{\sigma, \Sigma^V_\sigma - E^V_\sigma\})$
is purely discrete spectrum.
In particular $H^V_\sigma$ has a ground state.
\qed

\subsection{Case $\s=0$}
\label{massless}
Next we prove the existence of ground state of $H^V$.
For $\sigma>0$, let $\gr \in \hhh $ be a normalized ground state of $H^V_\sigma$.
Let $\{\sigma_n\}$ be a sequence such that $\lim_{n\to\infty} \sigma_n = 0$ and
$\Phi_{\sigma_n}$ converges weakly to some vector $\Phi\in\hhh $.
It is well known in \cite{ah97} that
if $\Phi\neq 0$ then $\Phi$ is a ground state of $H^V$.
In the following we prove that a subsequence of $\{\gr \}_\sigma$ converges to some
non-zero vector $\Phi$.
\begin{lemma} \label{h0bound}
 The energy bound $ \sup_{0<\sigma \ll 1} \inner{\gr }{ H_0 \gr } < \infty$ holds.
In addition we suppose $E^V<\Sigma^V$. Then
$ \sup_{0<\sigma \ll 1} \inner{\gr }{ N \gr }  < \infty$.
\end{lemma}
\proof
The former follows from the definition of $\gr$, and the
 later from \cite[Lemma IV2]{ger00}.
\qed
We denote the set of bounded operator on a Hilbert space $\cK$ by $B(\cK)$.
For each $k \in\BR$, let
$$\d v(k) = \sum_{j=1}^N\frac{\alpha_j}{\sqrt{2}} \la _j(-k) e^{-ikx_j}.$$
Then $v(k)\in B( L^2(\RR^{dN}_{X}))$.
For each $k\in\BR$, we set
\begin{align*}
 T(k) = (H^V -E^V +\ome(k))^{-1}  (v(k) \otimes\one_\fff) .
\end{align*}
Then $T(k)\in B(\hhh )$ for every $k\in\BR$, $\inner{\Psi}{T(k)\Phi}$ is measurable
for all $\Phi, \Psi \in \hhh $, and
$ \int_{\BR}\norm{T(k)}_{B(\hhh )}^2 dk < \infty$.
Hence $T(\cdot)$ can be regarded as a vector
in the Banach space $L^2(\BR;B(\hhh ))$.
Since $\gr\in D(N^\han)$, $a(k)\gr$ is well defined for almost every $k\in\BR$.
Let $\theta_s$, $s\in\BR$, be the shift on
$L^2(\BR; B(\hhh ))$, i.e., for $B\in L^2(\BR;B(\hhh ))$,
\begin{align*}
 (\theta_sB)(k) = B(k-s) , \qquad \mathrm{a.e.} k \in \BR.
\end{align*}
\begin{lemma} \label{lem 312}
 The map $\BR \ni s\mapsto \norm{\theta_sTe^{-\delta |x|}}_{L^2(\BR;B(\hhh ))}\in\RR$ is continuous.
\end{lemma}
\proof
Since $\theta_s$ is a translation, it is enough to show that
 $\|\theta_sTe^{\delta |x|}\|$ is continuous at $s=0$, i.e.,
$ \norm{ \theta_sTe^{-\delta |x|} - Te^{-\delta |x|} }_{L^2(\BR;B(\hhh ))} $
converges to 0 as $s\to 0$.
We have
\begin{align}
&  \norm{ \theta_sTe^{-\delta |x|} - Te^{-\delta |x|} }_{L^2(\BR;B(\hhh ))} \label{3120}\\
&   \leq
\lk
 \int_{|k|\leq C_1} +
 \int_{|k|\geq C_2} +
 \int_{ C_1 < |k| < C_2}
 \rk  \norm{ T(k-s)e^{-\delta |x|} - T(k)e^{-\delta |x|} }^2_{B(\hhh )} \notag
\end{align}
for $0<C_1<C_2$.
For $C_1<|k|<C_2$, we write
\begin{align*}
& T(k-s)e^{-\delta |x|} - T(k)e^{-\delta |x|} \\
& =
 (H^V - E^V + \ome(k))^{-1} \Big(\sum_{j=1}^N \Omega_j+\one\Big)
  \Big(\sum_{j=1}^N\Omega_j+\one\Big)^{-1}  (v(k-s)-v(k)) e^{-\delta |x|}\\
&\quad  + (H^V-E^V +\ome(k))^{-1}(H^V-E^V+\ome(k-s))^{-1}
  v(k-s)(\ome(k-s)-\ome(k))e^{-\delta |x|}.
\end{align*}
Since for all $k$ with $C_1<|k|<C_2$
\begin{align*}
 \sup_{C_1\leq |k|} \norm{(H^V-E^V+\ome(k))^{-1}\Big(\sum_{j=1}^N\Omega_j+\one\Big)}<\infty,
\end{align*}
 we have
\begin{align*}
& \norm{T(k-s)e^{-\delta |x|} - T(k)e^{-\delta |x|}}_{B(\hhh)} \\
& \leq C \norm{ \Big(\sum_{j=1}^N\Omega_j+\one\Big)^{-1}
e^{-\delta |x|}(v(k-s)-v(k))}_{B(\hhh)} +C \norm{ 
e^{-\delta |x|} v(k-s) }_{B(\hhh)}
\end{align*}
for some constant $C>0$ depending on $C_1$ and $C_2$.
Note that
$\Big(\sum_{j=1}^N\Omega_j+\one\Big)^{-1}e^{-\delta |x|}$ is compact.
By Proposition \ref{ge3.2} below,
we have
\begin{align}
 \lim_{s\to 0} \int_{C_1<|k|<C_2}
\norm{ \Big(\sum_{j=1}^N\Omega_j+\one\Big)^{-1}e^{-\delta |x|} (v(k-s)-v(k)) }_{B(\hhh)}^2 dk =0.
 \label{3121}
\end{align}
Next we see that
\begin{align*}
&\lim_{s\to 0} \int_{|k|\leq C_1} \norm{T(k-s)e^{-\delta |x|}-T(k)e^{-\delta |x|}}^2_{B(\hhh )} dk \\
& \leq 2\lim_{s\to 0}\int_{|k|\leq C_1} \left(\frac{|\la (-k)|^2}{|\ome(k)|^2}
 + \frac{|\la (-k+s)|^2}{|\ome(-k+s)|^2}\right)dk
 \leq
 4\int_{k\leq C_1}\frac{|\la (-k)|^2}{\ome(k)^2} dk,
  \end{align*}
and the right-hand side above converges to zero as
$ C_1 \to 0$.
Similarly,
\begin{align}
 \lim_{C_2\to\infty} \lim_{s\to 0}
 \int_{|k|\geq C_2} \norm{T(k-s)e^{-\delta |x|}-T(k)e^{-\delta |x|}}^2_{B(\hhh )} dk =0. \label{3123}
\end{align}
Therefore, by combining \eqref{3121} -- \eqref{3123}, we complete the proof.
\qed
\begin{proposition}
{\rm \cite[proof of Lemma 3.2]{ger06}}
{\label{ge3.2}}
 Let $\BR\ni k \mapsto m(k) \in B(L^2(\RR^{dN}))$ be a weakly measurable map
such that for all $0<C_1<C_2$,
\begin{align*}
  \int_{C_1\leq |k|\leq C_2} \norm{m(k)}_{B(L^2(\RR^{dN}))}^2 dk <\infty,
\end{align*}
and $R$ be a compact operator on $L^2(\RR^{dN})$. Then for
all $0<C_1<C_2$,
\begin{align*}
 \lim_{s\to 0} \int_{C_1<|k|<C_2} \norm{R(m(k-s)-m(k))}_{B(L^2(\RR^{dN}))}^2 dk =0.
\end{align*}
\end{proposition}
\begin{lemma} \label{bpb}
 Let $F \in C_0^\infty(\BR)$ be a cutoff function with
 $0\leq F \leq 1$, $F(s)=1$ for $|s|\leq 1/2$,
$F(s)=0$ for $|s|\geq 1$. Let $F_R=F_R(-i\nabla_k) = F(-i\nabla_k/R)$. Then
\begin{align}
 \lim_{R\to\infty} \sup_{0<\sigma \ll 1} \inner{\gr }{\dG (\one-F_R)\gr }
 =0 \label{318}
\end{align}
\end{lemma}
\proof
It is shown in \cite[proof of Proposition IV.3]{ger00} that
$$\d \lim_{\sigma\to 0} \int_{\BR}\norm{a(k)\gr - T(k)\gr }_\hhh ^2 dk = 0.$$
Then
\begin{align*}
 &\inner{\gr }{\dG (\one-F_R)\gr }_\hhh
 = \int_{\BR}
 \inner{T(k)\gr }{(\one -F_R) T(k) \gr }_\hhh
 dk  + o(\sigma^0),
\end{align*}
where $o(\sigma^0)$ denotes a constant converges to 0 as $\sigma\to 0$.
By Cauchy-Schwarz inequality yields that
the right-hand side above has the upper bound by
\begin{align} \label{sa7}
 \norm{T}_{L^2(\BR;B(\hhh ))} \cdot
 \norm{(\one-F_R)T(k)e^{-\delta |x|}}_{L^2(\RR_k^d;B(\hhh ))}
 \cdot
 \norm{e^{\delta |x|}\gr }_\hhh + o(\sigma^0)
\end{align}
Note that
$\sup_{0<\sigma \ll 1}\|e^{\delta |x|}\gr \|_\hhh <\infty$
for some $\delta>0$ by assumption.
By the Fourier transformation, we have
\begin{align}
& \norm{(\one-F_R) T(k) e^{-\delta |x|} }^2_{L^2(\BR;B(\hhh ))} \label{319}\\
& =
 \int_{\BR}\norm{
 (2\pi)^{-d/2} \int_\BR ds \hat{F}(s) (\one-\theta_{-s/R})T(k)e^{-\delta |x|}}^2_{B(\hhh )} dk \notag \\
& \leq
 (2\pi)^{-d}
 \int_{\BR} |\hat{F}(s)|^2 \cdot
 \norm{ (\one-\theta_{-s/R}) T e^{-\delta |x|}}_{L^2(\BR;B(\hhh ))} ds . \notag
\end{align}
Notice that
\begin{align*}
 |\hat{F}(s)|^2 \cdot \norm{(\one-\theta_{-s/R})Te^{-\delta |x|}}_{L^2(\BR;B(\hhh ))}
 \leq
 |\hat{F}(s)|^2 \cdot 2 \|\la _j/\ome\|,
\end{align*}
and the right-hand-side above is integrable in $s$ and independent of $R$.
Moreover, Lemma \ref{lem 312} implies that the integrand of the last
term in \eqref{319} converges to 0 as $R\to \infty$.
Therefore, by the Lebesgue dominated convergence theorem,
\eqref{319} converges to 0 as $R\to \infty$, and hence \eqref{318} holds.
 \qed

\begin{proposition} \label{sasaki5}{\bf (Proof of Proposition \ref{P3.1})}
$H^V$ has a ground state.
\end{proposition}
\proof
The proof is parallel with that of \cite[Lemma IV.5]{ger00}.
By $(\one-\Gamma(F_R))^2 \leq \dG (\one-F_R)$ and Lemma \ref{bpb},
we have
\begin{align}
 \norm{(\one-\Gamma(F_R))\gr } \leq o(R^0) + o(\sigma^0). \label{321}
\end{align}

Let $\{\sigma_n\}_n$ be the subsequence such that $\lim_{n\to\infty}\sigma_n = 0$
and $\Phi=\text{w-lim}_{n\to\infty}\Phi_{\sigma_n}$.
By Lemmas \ref{h0bound} and \ref{expdecay},
\eqref{321}, for all $\vep>0$, there exist $R_0>0$, 
$\lambda_0>0$, $n_0>0$ such that
for all $R>R_0$, $\lambda_0>\lambda$ and $n\geq n_0$,
\begin{align*}
& \norm{(\one-\chi(H_0\leq \lambda))\Phi_{\sigma_n}} <\vep, \qquad
 \norm{(\one-\chi(N \leq \lambda))\Phi_{\sigma_n}} <\vep, \\
&  \norm{(\one-\chi(|X| \leq \lambda))\Phi_{\sigma_n}} <\vep, \qquad
  \norm{(\one-\Gamma(F_R))\Phi_{\sigma_n}} <\vep,
\end{align*}
where $\chi(s\leq \lambda)$ denotes a characteristic function of support $\{s \in\RR| s< \lambda\}$.
Note that $K=\chi(H_0\leq \lambda)\chi(N\leq \lambda)\chi(|X|\leq\lambda)\Gamma(F_R)$
is a compact operator. For all large $R>0$, $\lambda>0$, we have
\begin{align*}
 \norm{\Phi}
&\geq \norm{K\Phi} - \norm{(\one-K)\Phi} \\
& \geq \lim_{ n \to \infty }\norm{K\Phi_{\sigma_n}} - \norm{(\one-K)\Phi} \\
& \geq \liminf_{ n \to \infty } (\norm{\Phi_{\sigma_n}}-\norm{(\one-K)\Phi_{\sigma_n}})
 - \norm{(\one-K)\Phi} \\
& \geq 1-4\vep - \norm{(\one-K)\Phi}.
\end{align*}
Clearly $\one-K$ strongly converges to $0$ when $R$ and $\lambda$ goes to infinity.
Since $\vep>0$ is arbitrary, we have $ \norm{\Phi}=1$.
Therefore $H^V$ has a normalized ground state $\Phi$.
\qed

\section{Essential spectrum}
\label{ess}
We give general lemmas given in \cite{hs08} without proofs.
\begin{lemma} \label{B1}
Let $K_\epsilon$, $\epsilon>0$, and $K$
 be self-adjoint operators on a Hilbert space ${\cal K}$
and
$\s_{\rm ess}(K_\epsilon)=[\xi_\epsilon,\infty)$.
Suppose that $\lim_{\epsilon\rightarrow 0}
K_\epsilon=K$ in the uniform
resolvent sense,
and $\lim_{\epsilon\rightarrow 0} \xi_\epsilon=\xi$.
Then $\s_{\rm ess}(K)=[\xi,\infty)$. In particular $\lim_{\e\rightarrow 0}\inf\!\sigma_{\rm ess}(K_\e)=\inf\!\sigma_{\rm ess}(K)$.
\end{lemma}
 \begin{lemma} \label{A1}
 Let
$\Delta$ be the $d$-dimensional Laplacian. Assume that
$V(-\Delta+1)^{-\han}$ is a compact operator. Then there exists a
sequence $\{V^\epsilon\}_{\epsilon>0}$ such that
$V^\epsilon\in
C_0^\infty(\BR)$ and $\lim_{\e\rightarrow 0} V^\epsilon(-\Delta 
+1)^{-\han }=V(-\Delta +1)^{-\han} $ uniformly. 
\end{lemma}
Set
$$
k_0(\beta)=-\sum_{j\in \beta}\sqrt{- \Delta_j}
+\sum_{i,j\in\beta} V_{ij},\quad k_V(\beta)=
h_0(\beta)+\sum_{j\in\beta}V_j
$$ 
with $V_i\in L_\mathrm{loc}^2(\BR)$ and $V_{ij} \in L_\mathrm{loc}^2(\BR)$ 
such that $V_i (-\Delta +1)^{-\han }$ and
$V_{ij}(-\Delta +1)^{-\han }$ are compact operators. 
We define $K=k_V(\N)$. Let
 \begin{align} \label{BB3}
 \Xi_V=\min_{\beta\subsetneqq C_N}
 \left \{\inf\!\sigma(k_0(\beta))+ \inf\!\sigma(k_V(\beta))\right \} 
 \end{align}
 be the lowest two cluster threshold of $K$.
\begin{lemma} \label{A2}
There exist sequences
$\{V_i^\e\}_\epsilon , \{V_{ij}^\e\}_\epsilon \subset C_0^\infty(\BR)$, $i,j=1,...,N$,
such that
$$ (1)\ \lim_{\epsilon\to0} \Xi_V(\epsilon)=\Xi_V,\quad
(2) \ \lim_{\epsilon\to0} \inf\!\sigma_{\rm ess}(K(\e))=\inf\!\sigma_{\rm ess}(K),$$
where $\Xi_V(\epsilon)$ (resp. $K(\e)$ ) is
$\Xi_V$ (resp. $K$)  with $V_i$ and $ V_{ij}$ replaced by
$V_i^\e$ and $V_{ij}^\e$, respectively.
\end{lemma}

\section{Functional integration and energy comparison inequality} \label{appb}
In this Appendix we shall show Lemma \ref{enebound1} and Proposition \ref{b2}
 by functional integrations.
In order to do that we take a Schr\"odinger representation instead of the Fock representation.
We quickly review the Schr\"odinger representation.

Let $\QQQ=\ms S_\RR'(\BR)$ be the set of real-valued Schwartz distributions on $\BR$.
The boson Fock space $\fff$ can be identified with $L^2(\QQQ,\mu)$ with some Gaussian measure $\mu$ such that
$\Ebb_\mu[\phi(f)]=0$ and $\Ebb_\mu[\phi(f)\phi(g)]=\half (f,g)$ for $f,g\in L_\RR^2(\BR)$.
Then the scalar field operator in $\fff$ is unitarily equivalent to the Gaussian 
random variable $\phi(f)$ in $L^2(\QQQ)$: 
$$\phi(f)\sim \frac{1}{\sqrt2}\int (a^{\ast}(k) \hat f(-k)+a(k) \hat f(k)) dk$$
for $f\in L_\RR^2(\BR)$. 
Moreover $\hf$ can be unitarily transformed to the self-adjoint operator in $L^2(\QQQ)$.
We denote it by the same notation, $\hf$.

Furthermore we need the Euclidean quantum field to construct the functional
 integral representation of the one-parameter semigroup generated by the Nelson Hamiltonian $H^V$.
Set $\QQQ_E=\ms S_\RR'(\RR^{d+1})$.
Thus $L^2(\QQQ_E,\mu_E)$ be the $L^2$ space endowed with a Gaussian measure such that
$\Ebb_{\mu_E}[\phi_E(F)]=0$ and $\Ebb_{\mu_E}[\phi_E(F)\phi_E(G)]=\half (F,G)_{L^2(\RR^{d+1})}$.
Let $j_t:L^2_\RR(\BR)\to L_\RR^2(\RR^{d+1})$ be the family of isometries connecting
$L^2(\QQQ)$ and $L^2(\QQQ_E)$, which satisfies that
 $j_s^\ast j_t=e^{-|t-s|\omega(-i\nabla)}$ for all $s,t\in\RR$.
Let $J_s=\Gamma(j_s)$ be the second quantization of $j_s$. Then $J_s:L^2(\QQQ)\to L^2(\QQQ_E)$ is
also the family of isometries such that $J_s^\ast J_t=e^{-|t-s|\hf}$ for all $s,t\in\RR$.
We identify $\hhh$ with the set of $L^2(\QQQ)$-valued $L^2$ function on $\BRN$, $\int^\oplus_\BRN L^2(\QQQ) dX$, and $H^V$
can be expressed as
\begin{align} \label{s10}\hp\otimes \one+\kappa^2 \one\otimes\hf+\kappa \alpha \sum_{j=1}^N \int^\oplus_{\BRN}
\phi(\lambda(\cdot-x_j)) dX
\end{align}
in the Schr\"odinger representation.

Next we prepare a probabilistic description of the kinetic term $\hp$.
Let $\pro X=(X_t^1,...,X_t^N)_{t\geq 0}$ be the $\BRN$-valued L\'evy processes on 
a probability space $({\cal D}, B, \PPPP^x)$ starting from $x=0$ 
with the characteristic function \eqref{char}.
Set $W(x_1,...,x_N)=\sum_{j=1}^N V(x_j)$.
 Then we have the Feynman-Kac formula:
 $$(f, e^{-\hp} g)=\int_\BRN\Ebb_\PPPP^x[\bar f(X_0)g(X_t)e^{-\int_0^t W(X_s) ds}].$$

The functional integral representation of $e^{-tH^V}$ can be obtained in the same way as
the standard Nelson model
Only the difference is the process associated with kinetic term.
Instead of the Brownian motion the L\'evy process $\pro {X^j}$ is entered for $e^{-tH^V}$.
The Feynman-Kac type formula of $e^{-tH^V}$ is then given by
\begin{align*}
& (F, e^{-tH^V}G)_\hhh = \\
&\int_{\BRN}
 dx \Ebb_\PPPP^x \left[ e^{-\int_0^t W(X_s) ds}
(J_0F(X_0), e^{-\kappa \phi_E ( \sum_{j=1}^N \int_0^t j_{\kappa^2 s}
\lambda_j (\cdot-X_s) ds ) } J_{\kappa^2 t} G(X_t) )_{L^2(\QQQ_E)} \right].
\end{align*}
Next we also consider the Feynman-Kac formula of $\exp(-t e^{-iT} H^V e^{iT})$.
It is
give in terms of the composition
of $dN$ dimensional Brownian motion $(B_t^1,...,B_t^N)_{t\geq0}$ on
a probability space
$({\cal C}, {\cal B}, \cW^x) $
and
$N$ independent subordinators $\pro {T^j}$, $j=1,...,N$, on
$(\Omega_\mu, {\cal B}_\mu, \mu)$ such that
$B_{T_t^j}^j$ has the same distribution of $X_t^j$.
Set $B_{T_t}=(B_{T^j_t}^j)_{t\geq 0,j=1,..,N}$.
We have the proposition below:
\begin{proposition} \label{f2}
Let $F,G\in \hhh$. Then
\begin{align*}
& (F, e^{-t e^{-iT} H^V e^{iT} }G)\\
&= e^{tE_\mathrm{diag}}
  \int_\BRN dx\Ebb_{\cW\times\mu}^{x,0}
  \left[e^{-\int_0^t (W+\eff)(B_{T_s}) ds}
 \lk
  J_0F(B_{T_0}), e^{-i\kappa^{-1}\phi_E
 \lk
  K_t
 \rk}
 J_{\kappa ^2 t}G(B_{T_t})\rk_{\!\! L^2(\QQQ_E)} \right].
\end{align*}
Here
$K_t=\sum_{j=1}^N
\int_0^{T_t^j}
 j_{({T^j}^{-1})_{\kappa^2 s}}
 \lambda_j(\cdot-B_s^j)\circ dB_s^j $ denotes the
 $L^2(\RR^{d+1})$-valued Stratonovich integral
 and $j_{({T^j}^{-1})_t}$ is some isometries defined by $\pro {T^j}$.
 \end{proposition}
\proof See \cite[Theorem 3.15]{Hir14}.
\qed

By using Proposition \ref{f2} we can compute the scaling limit of $e^{-iT} H^V e^{iT}$ as $\kappa\to \infty$.
Note that $(J_0\Phi, J_{\kappa^2 t}\Psi)\to (\Phi, P_\Omega \Phi)$ as $\kappa\to \infty$ for $t\not=0$.
Then by the functional integral representation Proposition \ref{f2} we immediately see that
\begin{align} \label{ball}
\lim_{\kappa\to\infty}(F, e^{-t e^{-iT} H^V e^{iT} }G)=
(F,
e^{-t
(\heff^V-E_\mathrm{diag})}
\otimes P_\Omega G).
\end{align}
Since $\heff^V$ has a ground state,
this suggests that $H^V$ also
has a ground state for sufficiently large $\kappa$. This has been indeed done in Section 3.

By functional integral representation we have the energy comparison bound.
\begin{proposition} \label{f3}
It follows that
$ \inf\!\sigma(H^V)\leq \inf\!\sigma(\heff^V)+E_\mathrm{diag}$.
\end{proposition}
\proof
By Proposition \ref{f2} we have
$|(F, e^{-t e^{-iT} H^V e^{iT} }G)|\leq
e^{tE_\mathrm{diag}}(|F|, e^{-t(\heff^V+\hf)}|G|)$.
Then the proposition follows.
\qed
In the same way as Proposition \ref{f3} but $H^V$ is replaced by $H^V(\beta)$ or $H^0(\beta)$ we have the lemma below.
\begin{proposition} \label{yui}{\bf (Lemma \ref{enebound1})}
It follows that
\begin{align} \label{koga}
\inf\!\sigma(H^\#(\beta)) \leq \inf\!\sigma(\heff^\#(\beta))+\sum_{j\in \beta}\frac{\alpha^2}{2}\|\la_j/\sqrt\omega\|^2,
\quad\#=0, V.
\end{align}
\end{proposition}
Next we show Proposition \ref{b2}.
We can also construct the functional integral representation of $e^{-tH^V_\s}$ in the quite same as that of $e^{-tH^V}$. Only the difference is to replace $\la_j$ with $\la_j\lceil_{\omega(k)>\s}$.
\begin{proposition} \label{ito}
Proposition \ref{b2} follows.
\end{proposition}
\proof
Notice that $\gr=e^{-t(e^{-iT} H_\s^Ve^{iT} -E_\s^V)}\gr$.
Then by Proposition \ref{f2} we can see that
\begin{align*}
\gr(x) = e^{t(E_\s^V+E_\mathrm{diag})}
\Ebb_{\cW\times\mu}^{x,0}
\left[e^{-\int_0^t \weff(B_{T_s}) ds}
J_0^\ast
e^{-i\kappa^{-1}
\phi_E \lk K_t
 \rk}
 J_{\kappa ^2 t}\gr(B_{T_t})\right].
\end{align*}
Thus it is straightforward to see by the Schwartz inequality that
\begin{align*}
\|\gr(x)\|_\fff\leq
e^{t(E_\s^V+E_\mathrm{diag})}
\lk
\Ebb_{\cW\times\mu}^{x,0}
\left[e^{-2\int_0^t \weff(B_{T_s}) ds}\right]\rk^\han
\|\gr\|_\hhh.
\end{align*}
Note that $\lim_{\s\to 0}E_\s^V=E^V$.
Then the proposition follows, since $B_{T_t}$ has the same distribution with $X_t$.
\qed

\section{Bound $E(0)\leq E(P)$ and continuity of $E(\cdot)$}
\label{appc}
 We consider a fiber decomposition of the translation invariant relativistic Schr\"odinger operator
$\hp=\sum_{j=1}^N \Omega_j+\eff$ in $\LRT$.

For notational convenience and generalizations, we consider the Schr\"odinger operator of the form
$\hp=\sum_{j=0}^N \Omega_j+v$ in $L^2(\RR^{d(N+1)})$, where $v=\sum_{j=0}^N v_{ij}(x_i-x_j)$ an we assume that $v$ is relativistic Kato-class.
Let $X_t=(X_t^j)_{t\geq 0}$, $j=0,...,N$, be
$N+1$ independent L\'evy processes with $\Ebb_\PPPP^x[e^{iu \cdot X_t^j}]=e^{-t \Omega_j(u)}$, and set
$\BX _t=(X_t^j)_{t\geq 0, j=1,...,N}$.
Let $\pt=\sum_{j=0}^N p_j $ be the total momentum. Then
$\hp$ commutes with $\pt$,
 and then
$
\hp\cong \int_\BR^\oplus k(P) dP$,
where $k(P)$ is a self-adjoint operator on $\LRT$. Let $E(P)=\inf\!\sigma(k(P))$.
\begin{theorem} \label{energy}
(1) $E(0)\leq E(P)$ for all $P\in\BR$.
(2) $\BR\ni P\mapsto E(P)\in \RR$ is continuous. 
\end{theorem}
We shall prove this theorem by making use of a path integral representation.
Let us set $x=(x_0,\bx)\in\BR\times \RR^{dN}$.
Let $U=F e^{ix_0 \cdot \ps }:L^2(\RR^{d(N+1)})\to L^2(\RR^{d(N+1)})$ be the unitary operator, where $F$ denotes the Fourier transformation with respect to $x_0$ variable, i.e.,
$Ff(k,\bx)=(2\pi)^{-d/2}\int f(x_0,\bx)e^{-ik \cdot x_0}dx_0$.
We have
$$(Uf)(k,\bx)=(2\pi)^{-d/2}\int_{\BR} e^{-ik \cdot x_0}f(x_0,x_1+x_0,\cdots,x_N+x_0)dx_0.$$
Thus we can directly see that
$(U\pt U^{-1} f)(k,\bx)=kf(k,\bx)$.
Hence $U$ diagonalize $\pt$, and thus
$ U\hp U^{-1} = \int_{\BR} k(P) dP$.
We have
\begin{align} \label{17}
(f, e^{-t\hp}g)_{L^2(\RR^{d(N+1)})}=\int_{\RR^{d(N+1)}}
 dx
 \Ebb_\PPPP^{(x_0,\bx)}
 \left[\ov{f(X_0)}g(X_t) e^{-\int_0^t v(X_s) ds}\right].
\end{align}
We construct the Feynman-Kac formula of $(f, e^{-tk(P)}g)_\LRT$.
Let $v=0$. Then
\[
 k(P)=\Omega_0\lk P-\ps\rk +\sum_{j=1}^N \Omega_j(p_j ).
\]
Since $\Ebb_\PPPP^{(0,\bx)} [e^{iX_t^0(P-\ps)}]=e^{-t\Omega_0(P-\ps)}$, we intuitively see that
$$
(f, e^{-tk(P)}g)_\LRT = \int_{\BRN} d\bx
\Ebb_\PPPP^{(0,\bx)}
[\ov{f(\BX _0)}e^{iX_t^0\cdot (P-\ps)}g(\BX _t)].
$$
Note that $e^{-iX_t^0\cdot \sum_{j=1}^N p_j }$ denotes a
translation,
 i.e.,
\begin{align*}
 (e^{-iX_t^0\cdot \sum_{j=1}^N p_j }g)(\BX_t) =
 g(X_t^1-X_t^0,\cdots, X_t^N-X_t^0).
\end{align*}
In the next proposition we see the Feynman-Kac formula
with potential.
\begin{proposition} \label{FKF2}
Let $F,G\in\LRT$ and $P\in\BR$.
Then
\begin{align} \label{18}
(F, e^{-tk(P)}G)_\LRT=
\int_{\RR^{dN}}d\bx
\Ebb_\PPPP^{(0,\bx)} \left[\ov{F(\BX_0)} e^{-\int_0^t v(X_s) ds}
e^{iX_t^0 \cdot (P-\ps)} G(\BX_t)
\right].
\end{align}
\end{proposition}
\proof
Let $\xi\in\BR$.
First we see that
\begin{align} \label{20}
(f, e^{-t\hp}e^{i\xi \cdot \pt}g)_{L^2(\RR^{d(N+1)})}=
\int_\BR dP e^{i\xi \cdot P}(f(P), e^{-tk(P)}g(P))_{\LRT},
\end{align}
where
$$f(P)=(Uf)(P,\bx)=(2\pi)^{-d/2}\int_{\BR} e^{-iP\cdot X}f(X,x_1+X,\cdots,x_N+X)dX,$$
and $g(P)$ is similarly given.
Now we put
$f=f_s=p_s\otimes F$ and
$g=g_r=p_r\otimes G$, where $F,G\in\ms S(\RR^{3N})$ and
$p_s(X)=(2\pi s)^{-d}\exp(-|X|^2/(2s))$ is the heat kernel.
Note that $f_s\to \delta(x_0)\otimes F$ as $s\downarrow0$.
We have
\begin{align*}
&\lim_{s\downarrow0} \int_{\BR} dPe^{i\xi \cdot P}(f_s(P), e^{-tk(P)}g_r(P))_\LRT \\
&=(2\pi)^{-d/2} \int_{\BR} dPe^{i\xi \cdot P} (F,e^{-tk(P)}g_r(P))_\LRT.
\end{align*}
The right hand side above is the inverse Fourier transform of the function
$h:P\to(F, e^{-tk(P)}g_r(P))_\LRT$ and
\begin{align} \label{na1}
\lim_{r\downarrow 0} h(P)=(F, e^{-k(P)}G)_\LRT(2\pi)^{-d/2}.
\end{align}
On the other hand
the left hand side of \eqref{20} can be represented by
the Feynman-Kac formula:
\begin{align} \label{21}
(f_s, e^{-t\hp}e^{i\xi \cdot \pt}g_r)
=
\int_{\mathbb{R}^{d(N+1)}} dx
\Ebb_\PPPP^{(x_0,\bx)}
\left[
\ov{f_s(X_0)}e^{-\int_0^t v(X_s) ds}
g_r(X_t^0+\xi , \cdots , X_t^N+\xi )
\right]
.
\end{align}
Taking $s\downarrow0$, we have
\begin{align*}
&
\int_{\mathbb{R}^{d(N+1)}} dx
\Ebb_\PPPP^{(x_0,\bx)}
\left[
\ov{f_s(X_0)}e^{-\int_0^t v(X_s) ds}
g_r(X_t^0+\xi,\cdots,X_t^N+\xi)
\right]\\
&\to
\Ebb_\PPPP^{(0,\bz)}
\left[\int_{\BRN} d\bx
\ov{F(\bx)}e^{-\int_0^t v(X_s+(0,\bx)) ds}
g_r(X_t^0+\xi,X_t^1+\xi+x_1\cdots,X_t^N+\xi+x_N)
\right].
\end{align*}
The right hand side is the function with respect to $\xi$.
We take the Fourier transform with respect to $\xi$. Then
\begin{align*}
\Ebb_\PPPP^{(0,\bz)}
\bigg[ & \int_{\BRN} d\bx
\ov{F(\bx)}e^{-\int_0^t v(X_s+(0,\bx)) ds} \\
 & \times (2\pi)^{-d/2} \int_{\BR} d\xi e^{-i\xi \cdot P}
g_r(X_t^0+\xi , X_t^1+\xi+x_1 , \cdots , X_t^N+\xi+x_N)
\bigg].
\end{align*}
Take $r\downarrow0$. We have
\begin{align*}
&\Ebb_\PPPP^{(0,\bz)}
\left[ \int_{\BRN} d\bx
  \ov{F(\bx)} e^{-\int_0^t v(X_s+(0,\bx)) ds} e^{iX_t^0 \cdot P}
G(X_t^1-X_t^0+x_1,\cdots,X_t^N-X_t^0+x_N)
\right]\\
=
&\Ebb_\PPPP^{(0,\bx)} \left[\int_{\BRN} d\bx
  \ov{F(X_0)}e^{-\int_0^t v(X_s) ds} e^{iX_t^0 (P-\ps)} G(\BX_t)
  \right].
\end{align*}
Comparing \eqref{na1} with the right hand side above,
 we conclude the theorem for $F,G \in\ms S$.
By a limiting argument the theorem is valid for all $f,g\in\LRT$.
\qed

{\it Proof of Theorem \ref{energy}:}
By Proposition \ref{FKF2} we have
\begin{align} \label{19}
|(f, e^{-tk(P)}g)| \leq
\int_{\RR^{dN}}dx\Ebb_\PPPP^{(0,\bx)}
\left[ |f(\BX_0)|  e^{-\int_0^t v(X_s) ds}  |e^{-iX_t^0 \cdot \ps} g(\BX_t)| \right].
\end{align}
Since $e^{-iX_t^0 \cdot \ps}$ is the shift operator,
$|e^{-iX_t^0 \cdot \ps} g(\BX_t)| \leq e^{-iX_t^0 \cdot \ps} |g(\BX_t)|$ follows.
 Then we obtain
$|(f, e^{-tk(P)}g)|\leq (|f|, e^{-tk(0)}|g|)$
which yields (1).

Next we show (2). By Feynman Kac formula
it is immediate to see that
\begin{align*}
& (F, (e^{-tk(P)}-e^{-tk(Q)})G)\\
&= \int_{\RR^{dN}}d\bx \, \Ebb_\PPPP^{(0,\bx)} \left[\ov{F(\BX_0)} e^{-\int_0^t v(X_s) ds}
e^{-iX_t^0 \cdot \ps}
\lk
i\int_{X_t^0 \cdot Q}^{X_t^0 \cdot P}e^{i\theta}d\theta\rk
 G(\BX_t)
\right].
\end{align*}
Then
$$\frac{|(F, (e^{-tk(P)}-e^{-tk(Q)})G)|}{\|F\|\|G\|}
\leq
|P-Q|
\sup_{\bx\in \BRN}
\lk
\Ebb_\PPPP^{(0,\bx)}
[|X_t^0|^2e^{-2\int_0^t v(X_s) ds}]\rk^\han.$$
Since $v$ is relativistic Kato-class,
\begin{align*}
\sup_{\bx\in \BRN}
\Ebb_\PPPP^{(0,\bx)}
[|X_t^0|^2e^{-2\int_0^t v(X_s) ds}]
\leq &
\sup_{\bx\in \BRN}  \Ebb_\PPPP^{(0,\bx)}  [|X_t^0|^4]^\han
\sup_{\bx\in \BRN}
  \lk  \Ebb_\PPPP^{(0,\bx)}[e^{-4\int_0^t v(X_s) ds}] \rk^\han \\
<&\infty.
\end{align*}
Then we conclude that $e^{-t k(P)}$ uniformly converges to $e^{-t k(Q)}$
as $|P-Q|\to 0$. Then (2) follows.
\qed

\section*{Acknowledgment}
FH acknowledges support of Grant-in-Aid for Science Research (B) 20340032 from JSPS,
Grant-in-Aid for Challenging Exploratory Research 22654018  and 
Grant-in-Aid for Challenging Exploratory Research 15K13445
from JSPS.
FH thanks a hospitality of ICMS Edinburgh, where this work is partially done.
IS acknowledges support of Grant-in-Aid for Young Scientists (B) 22740087 from JSPS.
IS thanks a hospitality of Orsay University, where this work is partially done.
IS's work is supported by the program for dissemination of tenure-track system 
funded by the ministry of education and science, Japan.

\bibliographystyle{alpha}
\bibliography{enhanced6.5}
\end{document}